\documentstyle[12pt,epsf]{article}

\begin{document}

\makeatletter
 \renewcommand{\theequation}{%
      \thesection.\arabic{equation}}
 \@addtoreset{equation}{section}
\makeatother

\begin{titlepage}
\begin{flushright}
EPHOU-02-003 \\
July, 2002
\end{flushright}
 
\vspace{15mm}
\begin{center} 
{\Large Numerical study for the $c$-dependence of fractal   \\}
{\Large dimension in two-dimensional quantum gravity}\\
\vspace{1cm}
{\bf Noboru Kawamoto
 \footnote{kawamoto@particle.sci.hokudai.ac.jp}
 and Kenji Yotsuji
 \footnote{present address: Visible Information Center, Inc. 440, 
 Muramatsu, Tokai-mura, Ibaraki, \\
\hspace*{5mm} Japan; yotsuji@masant.tokai.jaeri.go.jp}}\\
{\it{ Department of Physics, Faculty of Science }}\\
{\it{ Hokkaido University }}\\
{\it{ Sapporo, 060-0810, Japan}}\\
\end{center}
\vspace{1cm}

\begin{abstract}
We numerically investigate the fractal structure of two-dimensional 
quantum gravity coupled to matter central charge $c$ for $-2 \leq c \leq 1$. 
We reformulate $Q$-state Potts model into the model which can be identified 
as a weighted percolation cluster model and can make continuous change of 
$Q$, which relates $c$, on the dynamically triangulated lattice. 
The $c$-dependence of the critical coupling is measured from the percolation 
probability and susceptibility. The $c$-dependence of the string 
susceptibility of the quantum surface is evaluated and has very good agreement 
with the theoretical predictions. 
The $c$-dependence of the fractal dimension based on the finite size scaling 
hypothesis is measured and has excellent agreement with 
one of the theoretical predictions previously 
proposed except for the region near $c\approx 1$. 

\end{abstract}

\end{titlepage}

\section{Introduction}

It could probably be one of the most annoying questions: 
''What is the observable of quantum gravity to check if the quantum 
theory of gravity is well defined ?"
In two-dimensional quantum gravity the answer to this question is 
ready: ''The fractal dimension of the quantum surface is the well-defined 
observable which can be analytically and numerically calculated."  

The importance of the fractal nature of two-dimensional quantum gravity was 
first recognized by KPZ \cite{KPZ} where the critical exponent is 
recognized to represent the fractal structure of the two-dimensional 
quantum surface.
It has, however, been recognized later that more direct observable to detect 
the quantum fractal nature of space time would be the Housdorff dimension, 
or fractal dimension \cite{DHK,KN}. 

The serious numerical study to measure the fractal nature of two-dimensional 
quantum surface was initiated by Agistein and Migdal \cite{AM}.
They have proceeded the direct measurements of the fractal dimension 
of two-dimensional quantum surface by proposing a recursive 
sampling argorithm for $c=0$ model but could not observe the fractal nature of 
the quantum gravity in two dimension. The size of the triangulation was not 
large enough to observe the fractal nature of the quantum surface of $c=0$ 
model.    
The first numerical confirmation of the fractal structure of two-dimensional 
quantum gravity was carried out by Kawamoto, Kazakov, Saeki and Watabiki 
by using the recursive sampling argorithm for $c=-2$ model\cite{KKSW}. 
In these numerical analyses the large size 
lattice triangulation (up to 5 million triangles) was necessary to confirm the 
fractal nature by the direct measurement of the fractal dimension. 
This numerical confirmation of the two-dimensional quantum space-time 
triggered wide varieties of numerical and analytic investigations of the 
fractal nature of two-dimensional quantum gravity. 

There are three analytic derivations of the $c$-dependence of Housdorff 
dimension or fractal dimension of two-dimensional quantum gravity coupled 
to matter central charge $c$, by Distler, Hlousek and Kawai\cite{DHK}, 
Kawai and 
Ninomiya\cite{KN} and later by Watabiki, Kawamoto and Saeki\cite{KSW}. 
It was, however, pointed out that the measured fractal dimension of 
$c=-2$ model is very close to the third formulae given by Watabiki et. al.
\cite{KSW}. 
In the meantime Kawai, Kawamoto, Mogami and Watabiki tried to understand 
the fractal nature of quantum gravity from the Matrix model point of view 
and succeeded to derive the transfer matrix of the quantum surface of 
two-dimensional pure gravity ($c=0$)\cite{KKMW}.
The formulation made it possible to derive the fractal dimension of 
pure gravity to be exactly $d_F=4$ which is consistent with the value of the 
first and third formulae. 
This analytic investigation of the random surface triggered further analytic 
and numerical investigations of two-dimensional gravity\cite{W,AW,AKW,OTY}.  

Baby universe idea was proposed 
and proved to be useful to calculate string susceptibility 
numerically\cite{JM,AJT}. 
Then finite size scaling hypothesis was proposed later by being inspired 
by the analytic derivation of the two-point function of pure gravity\cite{AJW}.
The finite size scaling hypothesis made it 
possible to measure the fractal dimension very accurately. 
Using this formulation we made systematic and the most reliable numerical 
measurement of the fractal dimension for $c=-2$ model \cite{KWY1,KWY2}. 
Then the very accurately measured fractal 
dimension of $c=-2$ model perfectly agreed with the theoretical value 
of the third formula, $d_F=3.56\pm0.04(\hbox{numerical})\simeq 3.561\cdots
(\hbox{theoretical})$. 

So far the numerical investigations of the fractal dimensions were carried 
out mainly for $c=-2$ and $c=0$ model and for several unitary series 
of conformal field theory\cite{AA} and several values in the region 
$c>1$\cite{AT}. 

Here in this article we investigate the systematic investigation on the 
$c$-dependence of the fractal dimension of two-dimensional quantum surface. 
For $c=-2,0$ analytic formulae by the help of matrix model was available 
while for the other continuous value of matter central charge it was not 
obvious how to formulate the models to be convenient for the numerical study 
of the fractal dimension. It has, however, been known that the continuos 
central charge dependence can be accommodated by $Q$-state Potts model on 
the flat 
lattice. Here in this paper we reformulate the $Q$-state Potts model 
into the model which is a generalization of percolation cluster model, 
weighted percolation cluster model\cite{P,FK}. 
We formulate the model on the dynamically triangulated lattice 
and extend to take non-integer value $Q$ which is a function of $c$. 
By this model we can investigate $c$-dependence of the fractal nature of the 
two-dimensional quantum gravity coupled to matter central charge $c$. 
There have already been several calculations of critical indices of 
Ising model, 
three-state Potts models, and large $Q$ values of $Q$-state Potts model 
coupled to quantum gravity\cite{JKPS,ATW}.  

This paper is organized as follows: In section 2 we summarize the analytic 
derivations of Hausdorff dimension and fractal dimension. In section 3 
we explain the equivalence of Potts model and a weighted percolation cluster 
model which we call generalized Potts model. We then provide the definition 
of the generalized Potts model on the dynamically triangulated lattice. 
In section 4 we explain the details of the Metropolis argolism of the 
generalized Potts model. In section 5 numerical results of the 
$c$-dependence of critical coupling constant, string susceptibility, and 
fractal dimension are shown. 
Conclusions and discussions are given in the last section.

\section{Analytic derivation of fractal dimension 
and previous numerical results}

The fractal nature of the two-dimensional quantum gravity was first 
recognized by KPZ\cite{KPZ}. 
It was, however, not clear what kind of fractal it meant in the beginning. 
Serious analytic study on the fractal dimension of quantum gravity has been 
given by Distler, Hlousek and Kawai\cite{DHK}, and Kawai and 
Ninomiya\cite{KN} and later by Watabiki, Kawamoto and Saeki\cite{KSW}. 
Here we summarize the analytic derivation of the fractal dimension. 

In the derivation of the fractal dimension of two-dimensional 
quantum gravity coupled with matter central charge $c$, 
we use the formulation of Liouville theory in particular the 
formulation of conformal gauge given by DDK\cite{DDK}. 
We first summarize the main results of the formulation.

Formally the continuum partition function for matter coupled to two 
dimensional gravity is given by
\begin{equation}
 \label{eq:2.52}
 Z(A) \ = \ \int {\cal D}g ~\ \delta
 \left(\int d^2x\sqrt g -A\right) ~\, Z_M[g],
\end{equation}
where $Z_M[g]$ is a matter part of the partition function with 
gravitational background and $A$ is the total area.

Reguralized counterpart of the above partition function by dynamical 
triangulation is 
\begin{equation}
 \label{eq:2.52-1}
 Z_{reg}(A) \ = \ \sum_{G} \ Z_M[G] \, \delta_{Na^2,A}~~
              \sim \  Z_M[G_0],
\end{equation}
where $N$ is the number of equilateral triangles and $a^2$ is the area of 
the triangle.
$G$ denotes a triangulation and $G_0$ is the typical triangulation which 
we select from the huge set of triangulations.
The last approximate equality in eq.(\ref{eq:2.52-1}) is valid up to the 
normalization factor and if the selection of the typical surface is carried out by a correct procedure.
Since the path integration of the metric is carried out after the selection 
of the typical surface, $G_0$ carries the information of the quantum 
fluctuation of space time effectively. 

Following David, Distler and Kawai (DDK)\cite{DDK}, we obtain the gauge fixed 
version of the two-dimensional gravity coupled to matter central charge $c$ 
with a conformal gauge $g_{\mu\nu} = \hat{g}_{\mu\nu}e^\phi$:
\begin{equation}
 \label{eq:2.53}
Z(A) \ = \  
\int {\cal D}_{\hat g} \phi 
 \Delta_{\rm{FP}} [ \hat g ]  Z_{\rm M} [ \hat g ] \,  
 \delta \left( \int d^2 x \sqrt{\hat g} e^{\alpha_{1} \phi} 
               -  A \right)\, \exp \left(\frac{c-25}{48 \pi} 
S_{\rm L} [ \phi , \hat g ] \right), 
\end{equation}
where $\Delta_{\rm{FP}}$ is the Fadeev Popov contribution. 
$S_{\rm L} [ \phi , \hat g ]$ is the Liouville term given by 
\begin{equation}
 \label{eq:2.54}
S_{\rm L} [ \phi , \hat g ] \ = \ 
\int  d^2 x \sqrt{\hat{g}} \, 
\left( \, \frac{1}{2} \, \hat{g}^{\mu \nu} \partial_\mu \phi 
     \partial_\nu \phi \, + \, \phi \hat{R} \, \right) ,   
\end{equation}
where we set the renormalized cosmological constant equal to 
zero for simplicity.
$\alpha_1$ appeared in eq. (\ref{eq:2.53}) is obtained from the 
following general formula:
\begin{equation}
\label{eq:2.55}
 \alpha_{n} \ = \ 
 \frac{25 - c \, - \, \sqrt{ (25 - c)(25 - c - 24n) } }{12}.  
\end{equation}
DDK have shown that the primary conformal field of weight $n-1$ can be 
made Weyl invariant at the quantum level with a quantum correction:
\begin{equation}
\label{eq:2.56}
 \int  d^2 x \sqrt{\hat{g}} \, e^{\alpha_n\phi} \Phi_n, 
\end{equation}
where $\Phi_n$ transforms as 
$\Phi_n\mid_{\hat{g}e^\sigma} = e^{(n-1)\sigma}\Phi_n\mid_{\hat{g}}$.
The term $e^{\alpha_1\phi}$ in the delta function of eq. (\ref{eq:2.53})
is needed to keep the world sheet volume to be $A$ at the quantum level.

Let us define an expectation value of an observable $O(g)$ by 
\begin{eqnarray}
\label{eq:2.57}
<O(g)>_A
 \ &=&  \ Z(A)^{-1} 
\int {\cal D}_{\hat g} \phi
\Delta_{\rm{FP}} [ \hat g ]  Z_{\rm M} [ \hat g ]
\delta \left( \int d^2x \sqrt{\hat g} \, e^{\alpha_{1} \phi} 
               \, - \, A \right)\, 
\nonumber\\
&&O(\hat g, \phi) \, 
\exp\left(  \frac{c - 25}{48 \pi}   
S_{\rm L} [ \phi , \hat g ] \right).
\nonumber\\
\end{eqnarray}

\subsection{Housdorff dimension and fractal dimension}

Here we define two types of critical exponents which specialize 
the fractal nature of the two dimensional random surface. 

Let us first define an intrinsic area $A(r)$ of the random surface 
as a function of the mean square average size of the world sheet $<r^2>$ 
as viewed in the embedding space. 
We define Hausdorff dimension $d_H$ as 
\begin{equation}
 \label{eq:2.50}
   A(r) = (\sqrt{<r^2>})^{d_H}. 
\end{equation}
It should be noted that this definition of Hausdorff dimension refers 
to the embedded space. 

Let us next define $N(r)$ as the number of lattice points or number of 
triangles inside $r$ steps from a marked site. 
A step on the original lattice is one link step of a triangle while a 
step on the dual lattice is one dual link (edge) step on the dual lattice. 
We define fractal dimension $d_F$ as 
\begin{equation}
 \label{eq:2.51}
   N(r) = r^{d_F}. 
\end{equation}
This definition of the fractal dimension characterizes the connectivity of 
the random surface, how the random surface is composed by the connection of 
triangles, and thus could be called connectivity dimension. 

Distler, Hlousek and Kawai\cite{DHK} evaluated the mean square size of 
the quantum surface embedded in a $D$ dimensional space by calculating the 
two-point Green's function of vertex operator
\begin{eqnarray}
\label{eq:2.58}
\frac{1}{D}<x^2>_A \ 
&=&  \ 2\left| \frac{\partial}{\partial k^2} 
ln\left\langle \int d^2x_1\sqrt{\hat g(x_1)}\int d^2x_2\sqrt{\hat g(x_2)}
e^{ik(X(x_1)-X(x_2))} \right\rangle_A \right|_{k=0} \nonumber \\
&=&  \ \frac{1}{A^2Z(A)}
\left\langle \int d^2x_1\sqrt{\hat g(x_1)}\int d^2x_2\sqrt{\hat g(x_2)}
(X(x_1)-X(x_2))^2 \right\rangle_A \nonumber \\
&=& \ \left(\frac{A}{A_0}\right)^{\mid \gamma_{s}\mid} 
                   \hspace{1 cm} (A\rightarrow \infty), 
\end{eqnarray}
where $\gamma_{s}$ is the string susceptibility given by 
\begin{equation}
\label{eq:2.59}
 \gamma_{s} \ = \ 
 \frac{D - 1 \, - \, \sqrt{ (25 - D)(1 - D ) } }{12}.  
\end{equation}
We can then obtain the first formula of Housdorff dimension as 
a function of $D$.
\begin{equation}
\label{eq:2.60}
 d_H^{(1)} \ = \ \frac{2}{\mid\gamma_{s}\mid},  
\end{equation}
where $D$ could later be identified as matter central charge $c$ in two 
dimension. 

Let us next consider a derivation of fractal dimension using fermion 
as a test particle in the gravitational background following by Kawai 
and Ninomiya\cite{KN}. 
The Lagrangian for matter fermion coupled to gravity can be given by 
\begin{equation}
\label{eq:2.61}
 L \ = \ -\frac{1}{16\pi}\sqrt{g}R \ + \ \Lambda \sqrt{g} 
      \ + \ e\bar{\psi} i \not{\! \! D} \psi 
      \ - \ m e  \bar{\psi} \psi, 
\end{equation}
where $\Lambda$ and $e$ are, respectively, cosmological constant 
and the determinant of the vielbein in $D$-dimensions. 
In $D= 2 + \epsilon$ dimensions gravitational quantum corrections can 
be evaluated by the $\epsilon$-expansion formulation\cite{W}.
Then the fermion mass term is expected to acquire anomalous dimension via 
wave function renormalization of the matter fermion.  

Here we try to identify the anomalous dimension of the fermion mass term 
by the use of DDK formulation for Liouville theory.
Under the scaling of the cosmological constant 
\begin{equation}
\label{eq:2.62}
 \Lambda \ \rightarrow \beta^{-D} \Lambda, 
\end{equation}
the change in the Lagrangian can be absorbed by the field redefinition 
\begin{equation}
\label{eq:2.63}
 g_{\mu\nu} \ \rightarrow \ \beta^2 g_{\mu\nu}.
\end{equation}
Since the vielbein transforms like square root of metric, 
the fermion kinetic term transforms as 
$e\bar{\psi} i \not{\! \! D} \psi  
\rightarrow \beta^{D-1}e\bar{\psi} i \not{\! \! D} \psi$. 
Then the scaling parameter can be absorbed by the field redefinition 
$\psi \rightarrow \beta^{(1-D)/2}\psi$.  
The fermion mass term changes as 
$m e  \bar{\psi} \psi \rightarrow \beta m e  \bar{\psi} \psi$, 
in particular $\bar{\psi} \psi \rightarrow \beta^{-1}\bar{\psi} \psi$ 
in $D=2$.

In two dimensions the fermion mass term $m\int d^2x e \bar{\psi} \psi$ 
is expected to have the form of eq. (\ref{eq:2.56}) after the introduction 
of the gravitational quantum correction. Since $\sqrt{g}$ and $e$ acquire 
the same scale change, 
$\bar{\psi} \psi$ can be identified as primary 
conformal field of $\Phi_{1/2}$ because of the same scale change: 
$\Phi_{1/2}\mid_{\hat{g}\beta^2}=\beta^{-1}\Phi_{1/2}\mid_{\hat{g}}$ 
and 
$\bar{\psi} \psi \rightarrow \beta^{-1}\bar{\psi} \psi$. 
Then the Weyl invariant fermion mass term with the quantum correction 
is given by 
\begin{equation}
\label{eq:2.64}
 m\int d^2x \, \hat{e} \, e^{\alpha_{1/2}\phi} \, \bar{\psi} \psi.
\end{equation}

We now consider the following quantum average of the fermion mass term:
\begin{equation}
\label{eq:2.65}
 m\left\langle\int d^2x \, \hat{e} \, e^{\alpha_{1/2}\phi} \, 
 \bar{\psi} \psi\right\rangle_{A}, 
\end{equation}
where the quantum average $<..>_A$ is defined in eq. (\ref{eq:2.57}).
Under the constant shift of the conformal field 
$\phi \rightarrow \phi - 2ln\beta/\alpha_1$, the quantum average should 
be unchanged and yet the following relation holds:
\begin{equation}
\label{eq:2.66}
 m\left\langle\int d^2x \, \hat{e} \, e^{\alpha_{1/2}\phi} \, 
 \bar{\psi} \psi\right\rangle_{A}
\, = \, 
 m\beta^{-2\frac{\alpha_{1/2}}{\alpha_1}}\left\langle\int d^2x \, \hat{e} \, 
 e^{\alpha_{1/2}\phi} \, \bar{\psi} \psi\right\rangle_{\beta^{2}A},
\end{equation}
where the following change of delta function is taken into account: \\
$ \delta \left( \int  d^2 x \sqrt{\hat g} \, e^{\alpha_{1} \phi} 
\, - \, A \right)\, \rightarrow \, 
\beta^2 \delta \left( \int  d^2 x \sqrt{\hat g} \, e^{\alpha_{1} \phi} \, 
- \, \beta^2 A \right)$.
This relation suggests that the theory with two different sets of 
parameters $(A,m)$ and 
$\left(\beta^{2}A,m\beta^{-2\frac{\alpha_{1/2}}{\alpha_1}}\right)$ 
are equivalent.
The two dimensional volume measured by the length scale of 
the fermion field leads to another definition of fractal dimension 
$d_F^{(2)}$ \cite{KN},
\begin{equation}
\label{eq:2.67}
 A \, \sim \, L^2 \, = \, 
 \left(L^{2\frac{\alpha_{1/2}}{\alpha_1}}\right)^{d_F^{(2)}},
\end{equation}
where 
\begin{equation}
\label{eq:2.68}
 d_F^{(2)}(c) \, = \, \frac{\alpha_1}{\alpha_{1/2}} \, = \, 
 2 \times { \sqrt{ 25 - c } \, + \, \sqrt{ 13 - c }  \over 
            \sqrt{ 25 - c } \, + \, \sqrt{  1 - c }         },
\end{equation}
with $\alpha_n$ given by the formula (\ref{eq:2.55}).

\subsection{Fractal dimension from diffusion equation of random walk}

Here we provide yet another derivation of the fractal dimension by using 
the solution of diffusion equation on the random surface. 

We define the laplacian on the dynamically triangulated lattice. 
We first define adjacency matrix $K_{ij}$ on the dynamically triangulated 
lattice. For a chosen typical surface $G_0$ we number the sites of the 
triangulated lattice. Then the ($i,j$) component of the adjacency matrix 
$K_{ij}$ is defined as: $K_{ij}=1$ if $i$-th site and $j$-th site are 
connected by a link, $K_{ij}=0$ if they are not connected by a link. 
It is interesting to note that ($n,n_0$) component of $(K^T)_{i,j}$ counts 
the number of possible random walks reaching from a marking site $n_0$ to a 
site $n$ after $T$ steps.   
The Laplacian defined on the dynamically triangulated lattice is given by   
\begin{equation}
\label{eq:2.69}
 \Delta_L \ = \ 1 \, - \, S,~~~~~ S_{ij} \ = \ {1\over q_j}K_{ij},
\end{equation}
where $q_j$ is called coordination number and denotes a number of links 
connected to the site $j$. $S_{ij}$ is thus a probability of one step 
random walk from the site $j$ to the neighboring site $i$.
The diffusion equation on a triangulated surface $G_0$ with $N$ triangles is 
given by 
\begin{equation}
\label{eq:2.70}
\partial_T \Psi_N^{(G_0)}(T;n,n_0) \ = \ \Delta_L(G_0) \, 
\Psi_N^{(G_0)}(T;n,n_0),
\end{equation}
where $\partial_T$ is a difference operator in $T$ and 
$\Psi_N^{(G_0)}(T;n,n_0)$ is a wave function of the diffusion equation 
and denotes the probability of finding the random walker at the site $n$ 
after $T$ steps from the starting site $n_0$.
A solution of the diffusion equation can be obtained as 
$\Psi_N^{(G_0)}(T;n,n_0) = e^{T\Delta_L(G_0)}(\delta_{n,n_0})$, 
where $(\delta_{n,n_0})$ is $N$-component vector with unit $n_0$ entry.

We now consider the continuum limit of this diffusion equation.
First of all we recover the lattice constant $a$.
In taking continuum limit, the total physical area $A=a_i^2N_i$ is fixed 
and $i\rightarrow \infty; a_i\rightarrow 0, N_i\rightarrow \infty$ is taken, 
where $N_i$ is 
the number of triangles and $a_i^2$ is the area of a triangle.
In each step of the limiting process we select a typical surface $G_i$ 
for the given number of triangles $N_i$, on which the lattice Laplacian 
$\Delta_L(G_i)$ of eq.(\ref{eq:2.69}) is defined. 
Now the lattice version of the diffusion equation (\ref{eq:2.70}) can be 
rewritten by
\begin{equation}
\label{eq:2.71}
{1 \over a_i^2} \{ \Psi_A^{(G_i)}(T+a_i^2;x,x_0) \, - \, 
                   \Psi_A^{(G_i)}(T;x,x_0) \} \ = \ 
{1 \over a_i^2} \Delta_L(G_i) \Psi_A^{(G_i)}(T;x,x_0) \},
\end{equation}
where the location of the site $x$ is measured with respect to the lattice 
constant $a_i$.
Thus we identify the dimension of $T$ as that of area: $dim[T] =  dim[A]$.
In the continuum limit the solution of the diffusion equation 
(\ref{eq:2.71}) is expected to approach the continuum wave function: 
$\Psi_A^{(G_i)}(T;x,x_0) \rightarrow \Psi_A^{(G_{\infty})}(T;x,x_0)$.
Numerically we approximate the limiting surface as the typical surface($G_0$) 
of the maximum size triangulation: $G_\infty \simeq G_0$.
As we have already noted in eq.(\ref{eq:2.52-1}), the metric integration 
is effectively carried out for the equation (\ref{eq:2.71}) since we have 
chosen a typical surface.
This means that the quantum effect is included for the wave function of 
eq.(\ref{eq:2.71}).
On the other hand the solution of the continuum counterpart of the diffusion 
equation: $\partial_\tau \Psi(\tau;x,x_0) = \Delta(g) \Psi(\tau;x,x_0)$ 
is still background metric dependent in general.
Furthermore the dimensions of $T$ and $\tau$ may not necessarily be equal.   

Let us now define the comeback probability of random walk on the triangulated 
lattice and relate it with the continuum expression of Liouville theory as 
follows:
\begin{eqnarray} 
\label{eq:2.72}
G(T) \ &\equiv& \ \Psi_A^{(G_0)}(T;x =  x_0) \  \nonumber \\
& \simeq & \left . \ \left\langle \int d^2x \sqrt{ g} \, 
\Psi(\tau;x =  x_0)\right\rangle_A \right / 
\left\langle \int d^2x \sqrt{ g}\right\rangle_A \ \nonumber \\
& = & \ \frac{1}{A}\left\langle \int d^2x \sqrt{ g} \, e^{\tau \Delta} \, 
\Psi(0;x = x_0)\right\rangle_A \  \nonumber \\
& \sim & \ \frac{1}{A} \ \sim \ \frac{1}{T}, 
\end{eqnarray}
where $<..>_A$ is the quantum average defined in eq.(\ref{eq:2.57}) and 
the last similarity relations are dimensional relations.
We should remind of the fact that the metric integration is effectively carried out since we have chosen the typical surface $G_0$ for the wave function of the comeback probability.
The initial wave function can be formally written as 
$\Psi(0;x =  x_0) = \lim_{x \rightarrow x_0}  
\delta_\epsilon (x-x_0)1/\sqrt g$, 
where the regularized delta function is needed such as: 
$\delta_\epsilon(x-x_0)=(1/\pi)\times\epsilon/((x-x_0)^2+\epsilon^2)$.

We next consider how to accommodate the Weyl invariance into the diffusion 
equation of random walk at the quantum level by using the formulation of 
Liouville theory.
Let us consider the following quantity by Liouville theory: 
\begin{eqnarray}
\label{eq:2.73} 
\lefteqn{\left\langle\int d^2x \sqrt g \, 
\Psi(\tau;x =  x_0)\right\rangle_A}  \nonumber \\
\ &=& \ \left\langle\int d^2x \sqrt g \, 
\Psi(0;x = x_0)\right\rangle_A \ + \  
           \tau \, \left\langle\int d^2x \sqrt g \, \Delta \, 
           \Psi(0;x=x_0)\right\rangle_A \ + \ 
           \cdots, \nonumber\\
\end{eqnarray}
where the solution of the diffusion equation is expanded by $\tau$.
\\
Taking a conformal gauge $g_{\mu\nu}(x) = \hat g_{\mu\nu}e^{\phi(x)}$ and 
introducing DDK arguments, we can rewrite the first and second terms of 
the eq.(\ref{eq:2.73} ) as
\begin{eqnarray}
\label{eq:2.74} 
\left\langle\int d^2x \sqrt g \, \Psi(0;x = x_0)\right\rangle_A \ &=& \  
\left\langle\int d^2x \sqrt {\hat g} \, 
\big[\frac{1}{\sqrt {\hat g}}
\delta_\epsilon (x-  x_0)\big]_{x=x_0}\right\rangle_A 
\ = \ 1 ,
\nonumber\\ 
\left\langle\int d^2x \sqrt g \, \Delta \, 
\Psi(0;x =  x_0)\right\rangle_A \ &=& \  
\left\langle\int d^2x \sqrt {\hat g} \, e^{\alpha_{-1}\phi} \, 
\big[\overrightarrow {\hat \Delta_x} \, 
\frac{1}{\sqrt {\hat g}}
\delta_\epsilon (x - x_0)\big]_{x=x_0}\right\rangle_A, 
\nonumber \\
\end{eqnarray}
where the term $e^{\alpha_{-1}\phi}$ is introduced to keep the Weyl 
invariance of the second line of eq.(\ref{eq:2.74}) in accordance with 
the arguments of eq. (\ref{eq:2.56} ). 
Here it should be noted that $\Delta \, \Psi(0;x =  x_0)$ 
in the second term of eq. (\ref{eq:2.74}) can be identified 
as a primary conformal field of $\Phi_{-1}$.

Similar to the treatment for the Weyl invariance of the fermion mass 
term of eq. (\ref{eq:2.66}), the quantum average 
of the comeback probability should be unchanged under the constant shift 
of the conformal field $\phi \rightarrow \phi - 2ln\beta/\alpha_1$.  
In particular the second term of eq.(\ref{eq:2.73}) should be unchanged 
and yet the following parameter change is generated:
\begin{equation}
\label{eq:2.75}
 \tau\left\langle\int d^2x \, \sqrt{\hat{g}} \, e^{\alpha_{-1}\phi} \, 
 \hat{\Delta}\Psi(0;x=x_0)\right\rangle_{A}
\, = \, 
\tau\beta^{-2\frac{\alpha_{-1}}{\alpha_1}}\left\langle\int d^2x \, 
\sqrt{\hat{g}} \, e^{\alpha_{-1}\phi} \, 
 \hat{\Delta}\Psi(0;x=x_0)\right\rangle_{\beta^2A}.
\end{equation}
This relation suggests that the theory with two different sets of 
parameters $(\tau,A)$ and 
$\left(\beta^{-2\frac{\alpha_{-1}}{\alpha_1}}\tau,\beta^{2}A\right)$ 
are equivalent.
Then we obtain the following dimensional relation:
\begin{equation}
\label{eq:2.76}
\hbox{\rm dim}{ \tau } \ = \  
\hbox{\rm dim}{  A^{ - \frac{ \alpha_{-1}}{\alpha_{1} } } } . 
\end{equation}

We now point out that the expectation value of the mean squared geodesic distance is evaluated by the standard continuum treatment  
\begin{eqnarray}
\label{eq:2.77}
\lefteqn{\int d^2x \sqrt g \, \big\{ r(x,x_0) \big\}^2 \, \Psi(\tau;x,x_0)}
\nonumber\\
 \ &=& \  
 \int d^2x \sqrt g \, \big\{ r(x,x_0) \big\}^2 \, \left(
 \frac{1}{\sqrt{\hat{g}}}\delta(x-x_0) \, +\,  \tau{\hat \Delta_x} \, 
\frac{1}{\sqrt {\hat g}}\delta (x - x_0) \, + \cdots \right) 
\nonumber\\
 \ &=& \  -\, 4\tau \ + \ O(\tau^2),
\end{eqnarray}
which is related to the quantum version of the mean squared geodesic 
distance in the small $\tau$ region 
\begin{eqnarray}
\label{eq:2.78}
\lefteqn{<r^2>} \ \nonumber\\
&\equiv& \ \sum_x \, \{r(x,x_0)\}^2 \, \Psi_A^{(G_0)}(T;x,x_0)
\nonumber\\ 
&\simeq& \left . \left\langle\int d^2x \sqrt g \, \int d^2x_0 
\sqrt g \, \{r(x,x_0)\}^2 \, 
\Psi(\tau;x,x_0)\right\rangle_A \right / \left\langle\int d^2x 
\sqrt g\right\rangle_A \ 
\nonumber\\ 
&\sim& \ \tau \ \sim \ A^{-\frac{\alpha_{-1}}{\alpha_1}} \ 
\sim \ T^{-\frac{\alpha_{-1}}{\alpha_1}}.
\end{eqnarray}
The last similarity relations are dimensional relations.
Here we give the third definition of fractal dimension 
\begin{eqnarray}
\label{eq:2.79}
A\, = \, \left(\sqrt{<r^2>}\right)^{d_F^{(3)}}
%
\end{eqnarray}
where 
\begin{equation}
\label{eq:2.80}
 d_F^{(3)}(c)  
 \ = \ 
 - \, 2 \, \frac{ \alpha_{1}}{\alpha_{-1} } 
 \ = \ 
 2 \times \frac{ \sqrt{ 25 - c } \, + \, \sqrt{ 49 - c }}
               {\sqrt{ 25 - c } \, + \, \sqrt{  1 - c }         } ,  
\end{equation}
with $\alpha_n$ given by eq. (\ref{eq:2.55}).

\subsection{Previous numerical results}

Serious numerical investigations of the fractal nature of two-dimensional 
quantum gravity ($c=0$ model) was initiated by Agistein and Migdal
for $c=0$ model\cite{AM}. 
They proposed recursive sampling algorithm which partially use an analytic 
formula for tree diagrams and rainbow diagrams to generate planar 
Feynman diagrams corresponding to different types of two-dimensional sphere.  
They could't, however, observe the fractal structure numerically for this 
model. 
The fractal nature of two-dimensional quantum gravity was numerically 
first confirmed by Kawamoto, Kazakov, Saeki and Watabiki 
by using the recursive sampling algorithm for $c=-2$ model\cite{KKSW}. 
It was recognized in this numerical analyses that large number of triangles 
is necessary to confirm the fractal nature by the parameterization of 
the fractal dimension given by (\ref{eq:2.51}). 
In fact they needed 5 million triangles to confirm the fractal nature. 
It was then recognized that the number of triangles were not large enough 
to measure the fractal dimension of $c=0$ model. 

The numerical fractal dimension of $c=-2$ model was given in this numerical 
analyses as $d_F\simeq 3.55$. After deriving the third formula of the 
fractal dimension $d_F^{(3)}(c) $ of eq.(\ref{eq:2.80}), we have recognized 
that the numerical value of the fractal dimension is close to the theoretical 
value of the third formula $d_F^{(3)}(-2)=(3+\sqrt{17})/2=3.561\cdots$. 

Except for the analytic derivation of the fractal dimension, one obtains 
the following analytic relations of comeback probability in (\ref{eq:2.72}) 
and mean squared geodesic distance in (\ref{eq:2.78}) from the diffusion 
equation:   
\begin{enumerate}
\item{(1)} 
$G(T)T \ \sim \ const.$,
\item{(2)} 
$<r^2>  \ = \ T^{{2}/{d_F^{(3)}(-2)}} \ \simeq \ T^{-0.56}$, 
\\ 
\end{enumerate}
which can be numerically checked. These relations were numerically 
confirmed for $c=-2$ model in \cite{KSW}.

An alternative analytic investigation of the fractal structure of 
the pure gravity ($c=0$) were carried out by deriving transfer matrix 
of random surface by Kawai, Kawamoto, Mogami and Watabiki. 
They found out the beautiful scaling function $\rho(L;D)$ which counts 
the number of boundaries whose boundary lengths lie between $L$ and $L+dL$ 
located at geodesic distance $D$ measured from a marked point.       
It is evaluated by taking the continuum limit from the transfer matrix and disk amplitude of dynamical triangulation. 
The functional form of $\rho(L;D)$ for $c=0$ model is given by
\begin{equation}
\label{eq:2.81}
        \rho(L;D)D^2 \ = \ {3\over 7 \sqrt{\pi}}(x^{-5/2} + {1\over2} x^{-3/2}
 + {14\over3} x^{1/2})
                         e^{-x},     
\end{equation}
where $x=L/D^2$ is a scaling parameter.
This quantity $\rho(L;D)D^2$ for $c=0$ model was measured numerically and 
had excellent agreement with the theoretical scaling function 
(\ref{eq:2.81})\cite{OTY}. 
One of the important result of this analysis is that the fractal dimension 
of the $c=0$ model turns out to be $d_F=4$ which is consistent with 
the first and third formulae. 

Based on these theoretical and numerical evidences the third formula 
of the fractal dimension would be the correct formula for the $c$-dependence 
of the fractal dimension. 
In order to clear up the situation we started serious systematic 
and very accurate numerical study 
of $c=-2$ model by using finite size scaling hypothesis\cite{KWY1,KWY2}. 
It was concluded that the measured fractal dimension from this analysis 
is $d_F= 3.56 \pm 0.04$ which is perfectly consistent with the theoretical 
value of the third formula.



\section{Generalized Potts model $\equiv$ Weighted percolation cluster model}
Potts model \cite{P} was defined as a generalization of the Ising
model \cite{Isi}. 
Fortuin and Kasteleyn \cite{FK} showed that the $Q$-state Potts model is
equivalent to a weighted percolation cluster model which we explain 
in this section. 
Their construction allows the Potts model to be generalized to 
nonintegral values of $Q$.
We may call this model as generalized Potts model or equivalently 
weighted percolation cluster model.
These models were originally formulated on the square lattice 
while we extend to formulate the model on a dynamically triangulated 
lattice.  

We define a planar $\varphi^3$-graph ${\cal G}$ dual to 
a triangulated lattice.
Let a graph ${\cal G}_N$ have $N$ vertices which are dual to triangles 
in the triangulated lattice.
With each vertex $i$, we associate a spin 
that can take $Q$ different
values $\sigma_i = 1,2,\cdots ,Q$. Two adjacent spins $\sigma_i$ and
$\sigma_j$ interact with interaction energy 
$-J\, \delta (\sigma_i, \sigma_j)$, where
\begin{equation}
 \label{eq:4.3}
   \delta (\sigma_i, \sigma_j) = \left\{
    \begin{array}{ll}
      1, & \quad \sigma_i = \sigma_j \\
      0, & \quad \sigma_i \ne \sigma_j ~.
    \end{array} \right.
\end{equation}
Thus the Hamiltonian is 
\begin{equation}
 \label{eq:4.4}
   H = - K \sum_{\langle i,j\rangle } \delta (\sigma_i, \sigma_j)\, ,
\end{equation}
where $K = J/k_B T$ can be reinterpreted as a coupling constant and 
the summation runs over all the pairs of nearest-neighbor vertices 
$\langle i,j\rangle$.
Then the partition function of this model is given by
\begin{eqnarray}
 \label{eq:4.5}
   Z_N (K;{\cal G}_N) =  
   \sum_{\{\sigma\} \, \mbox{{\scriptsize on}} \,{\cal G}_N} 
   \exp\biggl[ K \sum_{\langle i,j\rangle }
   \delta (\sigma_i, \sigma_j)\biggr]\, ,
\end{eqnarray}
where the $\sigma$-summation runs over all possible values 
$\{\sigma_i = 1,2,\cdots ,Q\}$ for the spin 
variables $\sigma_1, \sigma_2, \cdots, \sigma_N$ on ${\cal G}_N$. 
Thus there are $Q^N$ terms in the summation.

It has been shown \cite{FK,BKW} that the partition function 
(\ref{eq:4.5}) can be expressed as a dichromatic polynomial \cite{Tut}.
In order to see this, let us expand the partition function as a 
product of terms associated with nearest-neighbor vertices. 
This can be worked out by 
using the relation $[\delta (\sigma_i, \sigma_j)]^m =
\delta (\sigma_i, \sigma_j) ~(m=1,2,\cdots)$ as follows
\begin{eqnarray}
 \label{eq:4.7}
   Z_N (K;{\cal G}_N) = 
   \sum_{\{\sigma\}} \prod_{\langle i,j\rangle }  
   \left[ 1 + v\, \delta (\sigma_i, \sigma_j)\right]\, ,
\end{eqnarray}
where we set $v = \exp(K) -1$.

Let $E$ be the number of the pairs of the nearest-neighbor vertices 
which we simply call edges on the graph ${\cal G}_N$. 
It is equal to the number of original links in the 
triangulated lattice, i.e. $3N/2$. 
Then the summand
in eq.(\ref{eq:4.7}) is a product of $E$ factors. Each factor is the sum
of two terms ($1$ and $v\, \delta (\sigma_i, \sigma_j)$), so the product
can be expanded as the sum of $2^E$ terms. Each of these $2^E$ terms can
be associated with a bond-cluster-graph (from now on call this a cluster
configuration) on ${\cal G}_N$. Note that the term is a product of 
$E$ factors, one for each edge. 
The factor for edge $\langle i,j\rangle$ is either $1$ or 
$v\, \delta (\sigma_i, \sigma_j)$: if 
it is the former, leave the edge empty, if the latter, place a bond on
the edge with weight $v\, \delta (\sigma_i, \sigma_j)$. Do this for all
edges $\langle i,j\rangle$. We then have a one-to-one correspondence between 
cluster configurations on ${\cal G}_N$ and terms in the expansion of the 
product in eq.(\ref{eq:4.7}).

\begin{figure}[t]
 \begin{center}
 \begin{minipage}{12cm}
   \epsfxsize=12cm \epsfbox {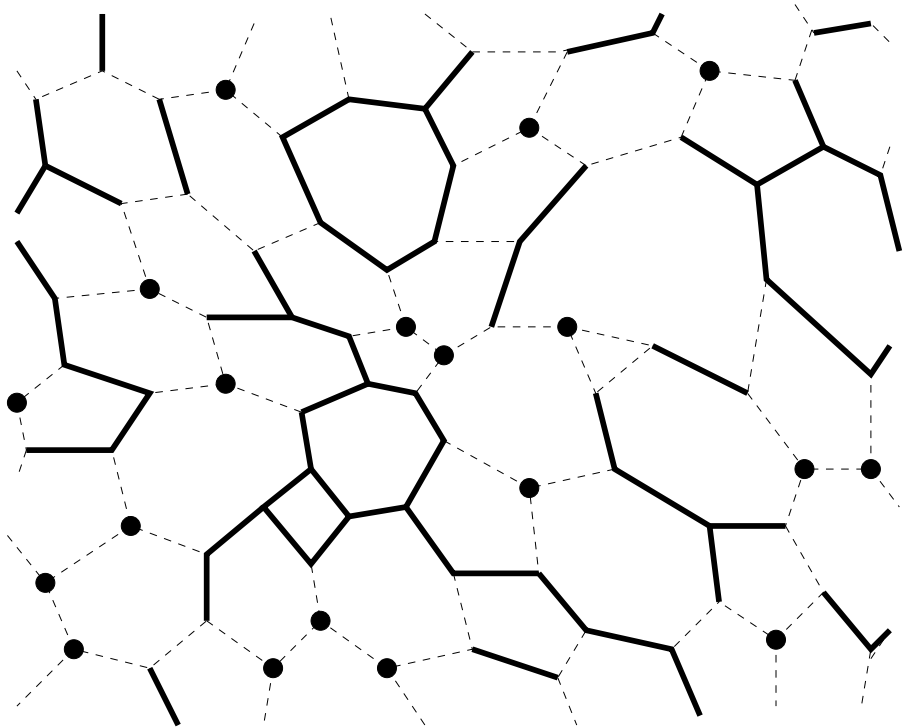}
 \end{minipage}
 \end{center}
 \caption{A fragment of a cluster configuration on a planar 
          $\varphi^3$-graph. An isolated vertex is regarded as a cluster.} 
 \label{fig:4.30}
\end{figure}

Consider a typical cluster configuration, 
containing $C$ connected components ($1\leq C\leq N$), namely, clusters and 
$b=\sum_{i=1}^C b_i$ bonds ($0\leq b\leq 3N/2$). 
See fig.\ref{fig:4.30}.
$i$-th cluster includes $b_i$ bands and an isolated vertex may be 
regarded as a cluster with $b_i=0$. 
Then the corresponding term in the expansion
contains a factor $v^b$, and the effect of the delta functions is that 
all vertices within a component must have the same spin $\sigma$. 
Summing over independent spins, it follows that terms gives a
contribution $Q^C v^b$ to the partition function $Z_N$. Summing over all
such terms, i.e. over all cluster configurations, we have
\begin{equation}
 \label{eq:4.8}
   Z_N^Q (K;{\cal G}_N) = 
   \sum_{{\scriptstyle \{cluster\} }
   \atop{\scriptstyle \mbox{{\scriptsize on}} ~{\cal G}_N}}  Q^C v^b\, .
\end{equation}
The summation is over all cluster configurations that can be drawn 
on ${\cal G}_N$. The expression (\ref{eq:4.8}) is a dichromatic 
polynomial. Note that $Q$ in eq.(\ref{eq:4.8}) need not be an integer.
We can allow it to be any positive real number, and this can be a 
useful generalization of $Q$-state Potts model to non-integer real number 
$Q$. We may call this model as {\it generalized $Q$-state Potts model} or 
{\it weighted percolation cluster model}. 

Let us consider eq.(\ref{eq:4.8}) for a few particular values of $Q$. 
Firstly, we consider a model $Z_N^{Q=1} (K)$ formulated on a given 
triangulated lattice. 
For the $Q\to 1$ limit, if we set $v=p/(1-p)$ the partition function is
\begin{eqnarray}
 \label{eq:4.9}
   Z_N^{Q=1} (K) & = & 
   \sum_{\{cluster\} } v^b \nonumber \\
   & = & \biggl( \frac{1}{1-p} \biggr)^E 
   \sum_{\{cluster\} } p^b (1-p)^{E-b}\, .
\end{eqnarray}
Therefore, $Z_N^{Q=1} (K)$ becomes a sum over all possible bond percolation
configurations with the correct weight where $p$ is the probability of a
bond being present \cite{FK}. This result holds in any dimension and 
any lattice on which one defines the Potts models. 
Since the sum $\sum_{\{cluster\} }$ in 
eq.(\ref{eq:4.9}) is the total probability and thus equal to $1$, 
this model coupled to two-dimensional quantum gravity corresponds to 
the pure gravity model ($c=0$). 

Next, let us examine for the $Q\to 0$ limit. At the critical point of 
the $Q$-state Potts model on the two-dimensional square lattice, 
it is known that $v\sim Q^{1/2}$ \cite{P}. 
In general, on a graph ${\cal G}_N$ we can assume $v~\sim Q^{\alpha}$ in
the $Q\to 0$ limit ($0 < \alpha < 1$). 
Then the partition function $Z_N^{Q\to 0} (K)$ becomes
\begin{equation}
 \label{eq:4.12}
   Z_N^{Q\to 0} (K) ~\sim ~Q^{\alpha N} 
   \sum_{\{ cluster\} } Q^{\alpha l + (1-\alpha ) C}\, ,
\end{equation}
where $l$ is the number of independent loops in a cluster configuration. 
We have used the Euler relation (See fig.\ref{fig:4.30}):
\begin{equation}
 \label{eq:4.13}
   b = N + l - C\, .
\end{equation}
For $0 < \alpha < 1$ the leading
terms in eq.(\ref{eq:4.12}) in the $Q\to 0$ limit can be obtained by 
taking $C=1$ (one cluster) and $l=0$ (no loops). 
These dominant configurations
are just the spanning trees of the graph ${\cal G}_N$ \cite{FK,Wu}.
Then each spanning tree configuration contributes to $Z_N^{Q\to 0} (K)$
with an equal weight. Therefore, this model coupled to 
two-dimensional quantum gravity is equivalent to the $c=-2$ scalar fermion
model coupled to quantum gravity \cite{KKSW}.

Temperley and Lieb \cite{TL} used the result of Fortuin and Kasteleyn
to prove the equivalence of the Potts model to the six-vertex or
square-ice model, with staggered polarizations. Baxter, Kelland and Wu
(BKW) \cite{BKW} have later found a very elegant derivation of the 
result of Temperley and Lieb. They use a construction known
as the BKW construction which makes many exact results obvious, including
the critical temperature, self-duality and energy at criticality of
the Potts model.

Thus the $Q$-state Potts models are analytically solved and the relations with 
the conformal field theories are well known \cite{Dot,DF}.
$Q$ is related to central charge $c$ in the following particular form: 
\begin{equation}
 \label{eq:4.1}
   Q = 4 \cos^2 \biggl(\frac{\pi}{m+1}\biggr), ~~~~~~~~
   c = 1 - \frac{6}{m(m+1)}\, .
\end{equation}
The minimal unitary conformal field theories with central
charge $c$ between $0$ and $1$ correspond to integer $m$; $m=2,3,4,\cdots$.
The generalization of $Q$ to any positive real number
corresponds to a continuous change of the central charge $c$, 
a generalization from minimal to non-minimal series of conformal 
field theories.

Within the framework of dynamical triangulations 
the generalized Potts model, or equivalently the weighted percolation
cluster models, coupled to two-dimensional quantum gravity is described by 
the following partition function:
\begin{equation}
 \label{eq:4.13.1}
   Z_N^Q (K) = 
   \sum_{{\cal G}_N\in \{\varphi^3 (T_N)\} }
   \frac{1}{{\cal S}_{{\cal G}_N}}\, 
   \sum_{\{cluster\} }  Q^C v^b \, .
\end{equation}
where $\{\varphi^3 (T_N)\}$ denotes the set of $\varphi^3 (T_N)$-graphs
dual to triangulations $T_N$ of fixed topology (which we always assume
to be sphere) and ${\cal S}_{{\cal G}_N}$ is a symmetry factor.

\section{The Metropolis algorithm of the generalized Potts model}
We intend to evaluate numerically the fractal dimensions of two dimensional 
quantum gravity coupled to matter central charge $c$ by the generalized 
Potts model, equivalently the weighted 
percolation cluster model formulated in the previous section 
(\ref{eq:4.13.1}). 
In the process of Metropolis updating we need double step updating: 
firstly we update a cluster configuration on a given graph ${\cal G}_N$, 
secondly the graph ${\cal G}_N$ itself should be updated. 

Here we first formulate the Metropolis algorithm of cluster configuration on a 
given graph ${\cal G}_N$.
The equation (\ref{eq:4.8}) leads us to choose the probability 
distribution of the generalized Potts model for a cluster configuration 
$\{C\}_k$ by 
\begin{equation}
 \label{eq:5.1}
   P[\{C\}_k]\, =\, \frac{Q^{C^{(k)}} v^{b^{(k)}}}{Z_N^Q (K;{\cal G}_N)}, 
\end{equation}
where $C^{(k)}$ and $b^{(k)}$ are the number of cluster and the total 
number of edges $b^{(k)}=\sum_{i=1}^{C^{(k)}} b^{(k)}_i$ 
for the given cluster configuration $\{C\}_k$, respectively.

We need to define a transition function $t[\{ C\}_k , \{ C\}_l] $ 
for a transition $\{ C\}_k \to \{ C\}_l$, 
which satisfies ergodicity and the following detailed balance condition: 
\begin{equation}
 \label{eq:5.2}
   P[\{ C\}_k ]\,\, t[\{ C\}_k , \{ C\}_l] =
   P[\{ C\}_l ]\,\, t[\{ C\}_l , \{ C\}_k] \, .
\end{equation}
Here we choose to use Glauber function \cite{Gla} as the transition function 
\begin{equation}
 \label{eq:5.3}
   t[\{ C\}_k , \{ C\}_l] =
   \, \frac{\delta W_{lk}}{1+\delta W_{lk}} \, ,
\end{equation}
where 
\begin{equation}
 \label{eq:4.21}
   \frac{t[\{ C\}_k , \{ C\}_l]}
        {t[\{ C\}_l , \{ C\}_k]} = \, \delta W_{lk} = \, 
        Q^{\delta C_{lk}}v^{\delta b_{lk}} 
\end{equation}
with $\delta C_{lk}=C^{(l)}-C^{(k)}$ and $\delta b_{lk}=b^{(l)}-b^{(k)}$.

For a given cluster configuration $\{C\}_k$, our updating proceeds as 
follows:
\begin{enumerate}
   \item[(1)]
   We randomly pick up an edge on the graph ${\cal G}_N$ and change the 
   edge by the following procedure and then the cluster 
   configuration changes into $\{C\}_l$.
   \item[(2)]
   We have to find out a change of the probability distribution 
   $\delta W_{lk} = Q^{\delta C_{lk}} v^{\delta b_{lk}}$  
   when we change the edge, where $\delta b_{lk}$ is the change in 
   the total number of bonds. If the edge originally has a bond, 
   remove the bond thus $\delta b_{lk} =-1$. 
   If the edge originally doesn't have a bond, add a bond to the edge thus 
   $\delta b_{lk} =+1$. 
   $\delta C_{lk}$ is the change in the number of clusters depending 
   on the corresponding change of the edge. 
   \item[(3)]
   Next, we generate a pseudorandom number $r$ uniformly distributed 
   from $0$ to $1$ and change the edge following to the procedure (2) if 
   and only if 
   \begin{equation}
    \label{eq:4.24}
      r \leq 
      \frac{\delta W_{lk}}{1+\delta W_{lk}} \, .
   \end{equation}
   This procedure ensures the transition $\{ C\}_k \to \{ C\}_l$ 
   with the correct probability.
   We use the Glauber function for the transition function 
   because of a faster 
   convergence to the equilibrium distribution in our model. 
   \item[(4)]
   Return to (1) unless the system is sufficiently equilibrated.
\end{enumerate}
By this updating, ergodicity and detailed balance condition can be
ensured.

\begin{figure}[t]
 \begin{center}
 \begin{minipage}{12cm}
   \epsfxsize=12cm \epsfbox {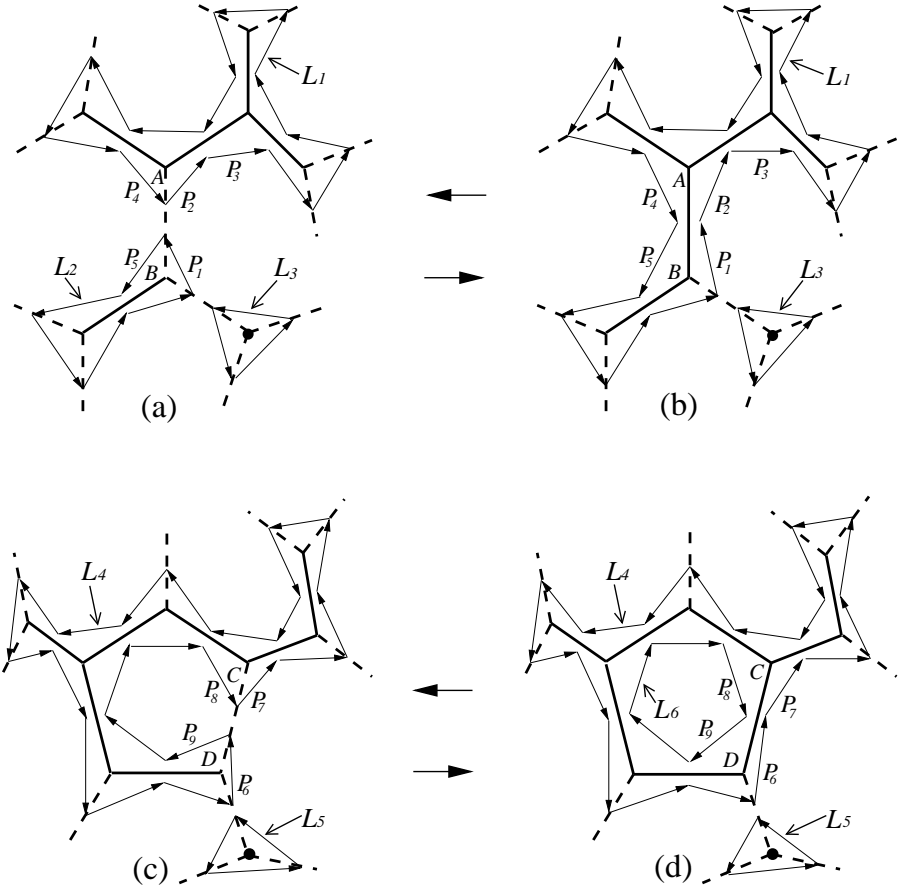}
 \end{minipage}
 \end{center}
 \caption{Cluster configurations: (a),(b),(c),(d).
          Doded lines are edges without bond while solid lines 
          are edges with bond. A and B are vertices in (a) and (b) 
          while C and D are vertices in (c) and (d). 
          $L_i$ denotes {\it i}-th surrounding loop. The arrows are 
          pointers and compose the segments of the surrounding loops. }
 \label{fig:4.37}
\end{figure}

Let us point out that it is nontrivial to evaluate 
$\delta C_{lk}$ in the step (2).  
For a given cluster configuration $\{C\}_k$, we pick up an 
edge on a graph ${\cal G}_N$. 
Suppose the edge does not have a bond 
we add a bond on the edge with the probability of the step (3) and 
thus $\delta b_{lk} =+1$ if the bond is added. 
For example we pick up the edge A-B which has vertices A and B in 
fig.\ref{fig:4.37}-(a) 
or the edge C-D which has vertices C and D in 
fig.\ref{fig:4.37}-(c), where those edges A-B 
and C-D does not have a bond. 
After the bond A-B and the bond C-D are added, 
(a) and (c) of fig.\ref{fig:4.37} turn into (b) and (d) of
fig.\ref{fig:4.37}, respectively.
In order to find $\delta C_{lk}$ we need to know if the both vertices of the 
edge belong to the same cluster or not before the bond is added. 
For example the vertices A and B in fig.\ref{fig:4.37}-(a) belong to 
the different clusters 
while the vertices C and D in fig.\ref{fig:4.37}-(c) belong to 
the same cluster. 
It is not time consuming to classify this difference numerically since 
we just need to know the data set of the collection of numbered 
vertices belonging to the same cluster. 
If both vertices originally belong to the different clusters then 
$\delta C_{lk}=-1$, 
while $\delta C_{lk}=0$ if they originally belong to the same cluster. 
It is thus numerically not difficult to identify $\delta C_{lk}$ in the 
case of $\delta b_{lk} =+1$. 

Let us suppose that we pick up an edge which already has a bond. Then we 
remove the bond with the probability of the step (3) and thus 
$\delta b_{lk} =-1$ if it is removed. 
We may pick up cluster configurations (b) and (d) of fig.\ref{fig:4.37} 
as particular examples which already has a bond at the edge A-B and 
C-D, respectively, and turn into (a) and (c) of fig.\ref{fig:4.37}, 
respectively, 
after the bond A-B and  C-D are removed.     
In order to find $\delta C_{lk}$ we need to know the information if the 
both vertices of the edge belong to the same cluster or not after the 
bond is removed. 
The vertices A and B belong to the different cluster in (a) and thus 
$\delta C_{lk}=+1$, while the vertices C and D belong to the same cluster 
in (c) and thus $\delta C_{lk}=0$. 
The crucial difference from the case $\delta b_{lk} =+1$ to the case 
$\delta b_{lk} =-1$ is that the collection of the data set to classify 
the different cluster is not enough to differentiate if two vertices 
are in the same cluster or not after the bond is removed.   
The straightforward way to determine 
the connectedness of two vertices is to start from the first vertex and
enumerate all vertices connected to it until either the second vertex is
reached, or an entire connected region will be enumerated. This method is
adequate if the clusters are small enough (tens of vertices), but if large 
clusters are involved (in the vicinity of the critical point) the
CPU time requirements can grow unreasonable. 
Since the connectedness, a non-local property, must be determined with every
iteration to know $\delta C_{lk}$, it is important to find 
faster algorithm to evaluate $\delta C_{lk}$. 
In order to quickly determine the connectedness of large clusters,
we have implemented an algorithm \cite{Swe} using an auxiliary data 
structure based on the BKW construction \cite{BKW}.
 
Let us reconsider 
a ${\varphi^3}$-graph ${\cal G}_N$, with $N$ vertices, together with its 
dual triangular lattice $T_N$. If a bond is present on ${\cal G}_N$, 
then its dual bond on $T_N$ is absent, and vice versa. The boundaries 
between clusters on ${\cal G}_N$ and their dual clusters on $T_N$ will 
form a collection of closed loops. Now we call this closed loops the 
surrounding loops. We then have the Euler relation,
\begin{equation}
 \label{eq:4.25}
   b + 2 C = N + L\, ,
\end{equation}
which relates the number of clusters and bonds to the number of surrounding
loops $L$. 

\begin{figure}[t]
 \begin{center}
 \begin{minipage}{4cm}
   \epsfxsize=4cm \epsfbox {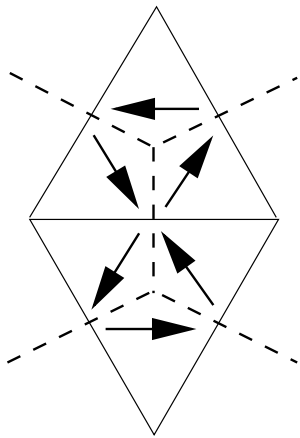}
 \end{minipage}
 \end{center}
 \caption{Three pointers surrounding each vertex.}
 \label{fig:4.35}
\end{figure}

By saving information of the surrounding loops as a data set, 
we transform the problem of determining connectedness of two vertices 
to the problem of determining whether 
two surrounding loop segments are part of the same surrounding loop or not.
The surrounding loops are represented as a chain of pointers in the computer. 
A pointer is a memory location containing the address of the next pointer 
in the chain.
There are three surrounding pointers represented by 
arrows for each vertex, as is shown in fig.\ref{fig:4.35}. 
We have shown surrounding loops composed of pointers for the figures 
(a), (b), (c), and (d) of fig.\ref{fig:4.37}.
For example in fig.\ref{fig:4.37}-(b) pointer $P_1$ points to 
pointer $P_2$, $P_2$ points to $P_3$, etc. In this manner the loops are
represented by chains of pointers. Because of the (differential) Euler 
relation, $\delta b_{lk} + 2\delta C_{lk} = \delta L_{lk}$, we can 
determine $\delta C_{lk}$
if we can determine $\delta L_{lk}$, the change in the number of surrounding
loops. 

Now, let us consider cases of removing a bond in the process of a 
Metropolis updating in Fig.\ref{fig:4.37}. The change; fig.\ref{fig:4.37}-(b) 
$\rightarrow$ fig.\ref{fig:4.37}-(a), 
illustrates the case in which removing the bond A-B will 
divide the surrounding loop $L_1$ into two surrounding loops $L_1$ and 
$L_2$ in fig.\ref{fig:4.37}-(a), while the change; fig.\ref{fig:4.37}-(d) 
$\rightarrow$ fig.\ref{fig:4.37}-(c), 
illustrates the case in which two surrounding loops $L_4$ and 
$L_6$ in fig.\ref{fig:4.37}-(d) will be joined into the loop 
$L_4$ in fig.\ref{fig:4.37}-(c) 
if the bond C-D is removed. 
In either case, two cuts must be made in chains of pointers and
the four ends rejoined if the bond is removed. In the case of A-B bond in 
fig.\ref{fig:4.37}-(b), the $P_1\to P_2$ and the $P_4\to P_5$ 
connections must be 
replaced by $P_1\to P_5$ and $P_4\to P_2$ connections, respectively, 
in fig.\ref{fig:4.37}-(a). 
Similarly in the case of the bond C-D in fig.\ref{fig:4.37}-(d), 
the $P_6\to P_7$ and 
the $P_8\to P_9$ connections must be replaced by $P_6\to P_9$ and 
$P_8\to P_7$ connections, respectively, in fig.\ref{fig:4.37}-(c). 
By collecting the information of chains we can immediately conclude 
that $\delta C_{lk}=+1 \, (\hbox{for}\, \delta b_{lk}=-1)$ since $P_1$ and 
$P_4$ are 
in the same chain for fig.\ref{fig:4.37}-(b) while $P_6$ and $P_8$ 
are in the different chains for fig.\ref{fig:4.37}-(d) and thus 
$\delta C_{lk}=0 \, (\hbox{for}\, \delta b_{lk}=-1)$. 
It is important to note that the imformation of the connectedness of the 
cluster configuration of (a) and (c) can be obtained by the data set 
of the chains of pointers of (b) and (d), respectively.  
It is numerically much easier to find if two pointers are in the same 
loop chain or not.  

In summary, in order to find $\delta W_{lk}$ for the Metropolis algorithm 
of the generalized Potts model in Monte Carlo simulation, we first pick up 
an edge and find out 
$\delta b_{lk}$ depending on whether the edge has a bond or not. 
We first cut and rejoin our chains of pointers depending on 
addition or removal of the bond on the edge. 
Using the (differential) Euler 
relation, $\delta b_{lk} + 2\delta C_{lk} = \delta L_{lk}$, we 
evaluate $\delta C_{lk}$ 
by judging whether two pointers attached to the different vertices of the 
given edge belong to the same chain or not.

\begin{figure}[t]
 \begin{center}
 \begin{minipage}{12cm}
   \epsfxsize=12cm \epsfbox {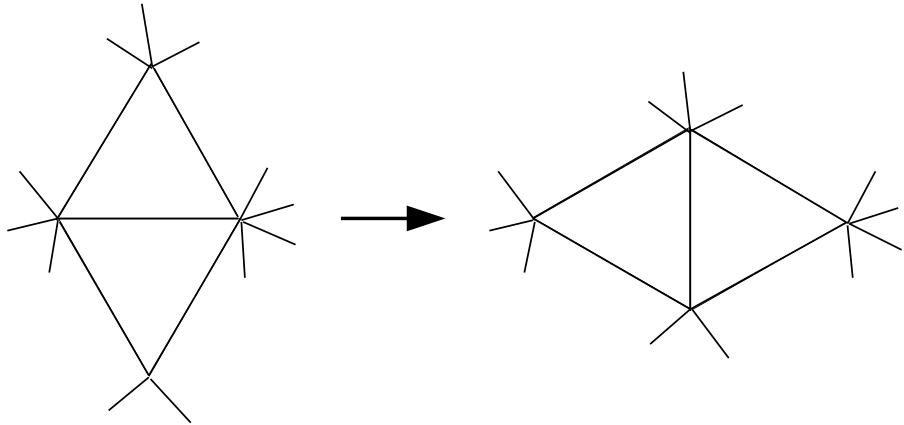}
 \end{minipage}
 \end{center}
 \caption{The flip move. This move is ergodic in the class of triangulations
          with a fixed number of triangles and fixed topology.}
 \label{fig:4.33}
\end{figure}

When we apply Monte Carlo simulations to quantum gravity coupled to the 
generalized Potts model or equivalently the weighted percolation clusters, 
we have to update the cluster configuration for a given triangulation and 
at the same time we have to update the triangulation for a given cluster 
configuration. 
The updating of cluster configurations on a given triangulation can be 
carried out just as described in the above.
In order to update triangulations corresponding to the given $\varphi^3$-
graphs ${\cal G}_N$, we use the standard flip-flop algorithm. 
We first choose two neighboring triangles randomly and flip the common link 
to generate a new triangulation. This flip move changes the triangulation 
locally as in fig.\ref{fig:4.33}. This move is enough to make the process 
ergodic for the chosen class of triangulations; fixed number of triangles and 
fixed topology \cite{BKKM}.

In our simulations we avoid to generate configurations corresponding to 
tadpole and self-energy graphs.
In other words we consider the class of triangulations 
satisfying the following conditions: 
\begin{enumerate}
   \item[(1)]
   No link has coinciding end.
   \item[(2)]
   Two sites may be connected by no more than one link (i.e. parallel 
   links are forbidden).
\end{enumerate}
In the updating of triangulations, if a triangulation in a forbidden class
is generated we ignore the attempt to make the flip and instead choose
a new link randomly and attempt a new flip. 

\begin{figure}[t]
 \begin{center}
 \begin{minipage}{12cm}
   \epsfxsize=12cm \epsfbox {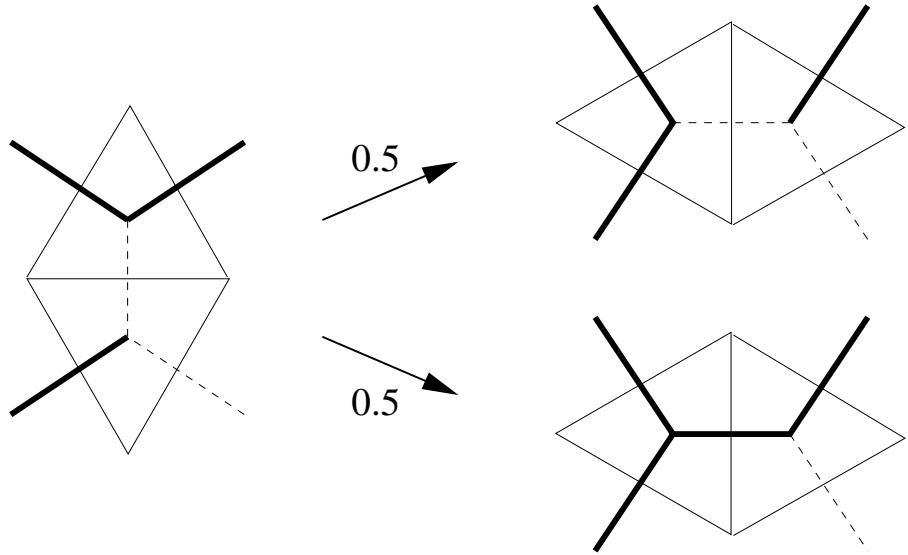}
 \end{minipage}
 \end{center}
 \caption{Bond assignment 
          after a flip move in a fragment of a cluster configuration.
          The bond is assigned to the new edge with 50 \% probability.}
 \label{fig:4.34}
\end{figure}

It is important to recognize that the flip procedure to generate random 
surface changes the cluster configuration. 
By the flip procedure the chosen edge which is the dual of the common 
link of the chosen neighboring triangles changes into a new 
edge which may or may not have a bond. 
In the present algorithm we generate a bond on the 
new edge with 50\% probability. For example a cluster configuration in 
fig.\ref{fig:4.34} changes into two possible cluster configurations 
with an equal probability by the flip procedure. 
In this way a change in
connectivity of two triangles concerned with updating entails a change in
the assignments of neighbors and therefore a change in the weight function
$  P[\{C\}_k]\, =\, Q^{C^{(k)}} v^{b^{(k)}}/Z_N^Q (K;{\cal G}_N)$ 
and then the transition function, $\delta W_{lk}/(1+\delta W_{lk})$, 
can be estimated according to the Metropolis scheme described in the above.

\section{Numerical simulations and results}
The Monte Carlo simulations of generalized Potts models on 
dynamically triangulated surfaces are performed using the algorithm 
mentioned in the previous section. In our simulations we only generate planar
$\varphi^3$-graphs of spherical topology with no tadpole or self-energy 
configurations. And we perform simulations on graphs with the fixed number $N$ 
vertices.

We first define one sweep of our Monte Carlo simulation. In the process of
updating for a given cluster configuration with a fixed triangulation 
one cluster-sweep 
means checking of the Metropolis procedure on each edge of the graph
in a given order. On the other hand, in the process of updating dynamical 
triangulation, one triangulation-sweep means that we 
randomly pick up edges one after another with roughly $3N/2$ (total number 
of edges) times and try to flip them and at the same time we check to 
eliminate every triangulations containing tadpole or self-energy graph.

After generating an initial cluster configuration, we thermalize
our system by performing 2000 triangulation-sweeps and 1000 cluster-sweeps
respectively. For the observables to be measured, this thermalization time 
is enough to thermalize our system according to the measurement of the 
autocorrelation time at criticality. To obtain a typical cluster sample 
configuration on a typical random surface background 
we perform 100 triangulation-sweeps and 20 cluster-sweeps. Then we measure 
several observables for the given configuration. We iterate these 
operations enough times until independent samples of a given number, 
which depends on the lattice size 
and on the observable to be measured, are obtained.

\subsection{The critical coupling constants}
It is a well accepted observation that lattice models with a second 
order phase transition lead to corresponding continuum theories 
at the second order phase transition point. In particular minimal conformal 
lattice models lead to the corresponding continuum conformal field theory 
models at the critical point in two dimensions. 
It is well known that the $Q$-state Potts models on regular two-dimensional 
lattice correspond to the field theories of minimal unitary conformal series 
at the corresponding critical coupling constant $K_c$. 
The correspondence is given in eq. (\ref{eq:4.1}).
It is analytically known that the $Q$-state Potts models make continuous 
(second order or even higher order) phase transition at the 
critical point for $0\leq Q\leq 4$, while they have first order
phase transition point for $Q > 4$. 

It is analytically not known if the nature of the phase transition 
may be changed when gravity coupled to the Potts models.  
For several examples of minimal unitary models, it is known that 
the order of phase transition can be raised to higher order when 
gravity is coupled \cite{Kaz,BK}.
In the case of $Q=10$ and $200$, numerical results 
suggests a strong evidence in favor of continuous transitions \cite{ATW}.
In our simulations we assume that the $Q$ state Potts models coupled to 
gravity presented by eq.(\ref{eq:4.13.1}) have continuous phase transitions 
for $0\leq Q\leq 4$ even at the non-unitary value of $c$.

In order to locate the critical coupling by using finite-size scaling method,
we investigate the percolation probability $P(K)$ and the cluster size
distribution $n_s (K)$ of percolation theory \cite{Sta}. 
Let us briefly summarize the physical meaning of the percolation 
probability $P(K)$ and the cluster size distribution $n_s (K)$ 
by studying the simplest Potts model of $Q=1$ case.
The partition function of $Q=1$ Potts model $Z_N^{Q=1} (K)$ is given 
by eq.(\ref{eq:4.9}). 
By the last expression of eq.(\ref{eq:4.9}) we can recognize that 
$p$ with $v=e^K-1=p/(1-p)$ can be identified as the probability of a
bond being present at an edge \cite{FK}.
When $K\rightarrow 0, v\rightarrow 0$ and thus $p\rightarrow 0$, $i.e.$ 
the probability of a bond being present at an edge is getting small, 
then the probability of finding a maximum cluster (the bonds of the 
maximum cluster reaches from one end to the other of the lattice 
extension) is expected to be zero. 
Let us define a quantity $P_N(K)={\hbox {\{the maximum cluster size\}}}/N$, 
where the cluster size is defined as a number of bond of the cluster. 
$P_N(K)$ is a probability of a vertex being on the maximum cluster, where 
$N$ is the total number of vertices. 
$P(K)=\lim_{N\rightarrow \infty}P_N(K)$ is called percolation probability. 
This quantity is
the order parameter of the percolation transition and is expected to show 
the following critical singularity in the infinite lattice 
for $|K-K_c|\to 0$  
\begin{equation}
 \label{eq:4.26}
   P(K) \sim \left\{
    \begin{array}{ll}
      (K-K_c)^{\beta_p}\, , & \quad K \geq K_c \\
      0\, , & \quad K \leq K_c ~,
    \end{array} \right.
\end{equation}
where $\beta_p$ is the critical exponent associated to the percolation 
probability.

\begin{figure}[t]
 \vspace*{-0.5cm}
 \begin{tabular}{cc}
 \begin{minipage}{7cm}
   \epsfxsize=7cm \epsfbox {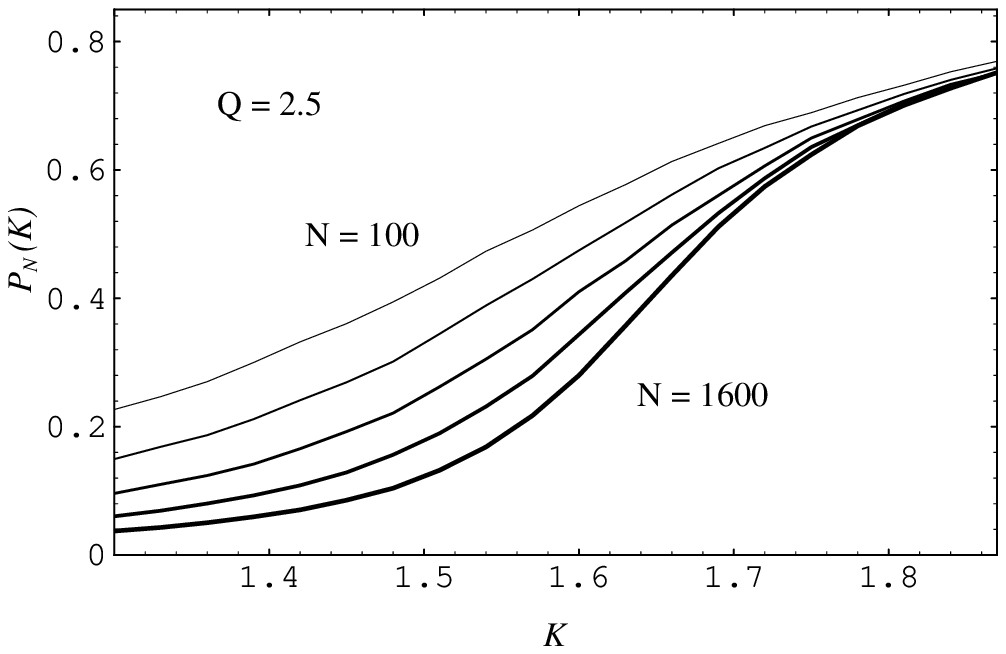}
 \end{minipage} &
 \begin{minipage}{7cm}
   \epsfxsize=7cm \epsfbox {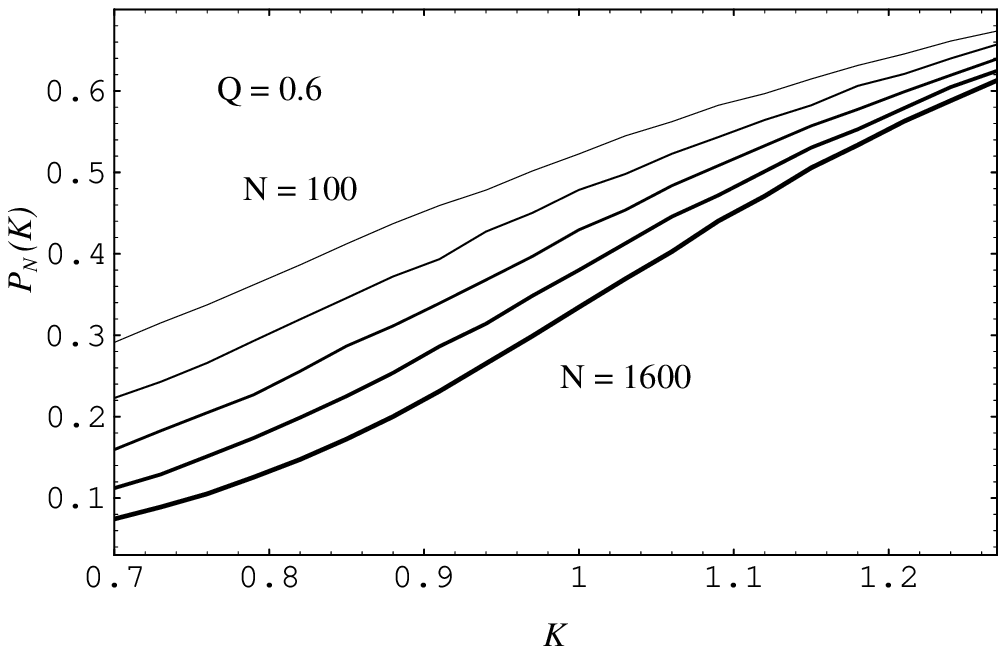}
 \end{minipage} 
 \end{tabular}
 \vspace*{-1.0cm}
 \caption{Size-dependence of $P_N(K)$ for $Q=2.5$ and $0.6$ as examples. 
          The sizes of the systems are $N=100$ (the highest
          curve), $200,\, 400,\, 800$ and $1600$ (the lowest curve).}
 \label{fig:4.2}
\end{figure}

In a finite lattice with the size $N$, the finite-size
percolation probability $P_N (K)$ can not vanish at any $K > 0$, and
its behavior depends on the lattice size. 
In fig.\ref{fig:4.2}  we show
the size dependence of $P_N (K)$ for $Q=2.5$ and $0.6$ as examples. 
As we can see, $P_N (K)$ with finite size dependence does not have sharp 
rise in contrast with eq.(\ref{eq:4.26}) but has milder rise with 
respect to $K$.
In these simulations
we have performed with the lattice sizes $N=100,\, 200,\, 400,\, 800$ and 
$1600$ for
$Q=0.2,\, 0.4,\, 0.6,\, 0.8,\, 1.0,\, 1.5,\, 2.0,\,$ 
$2.5,\, 3.0,\, 3.5$ 
and $4.0$ respectively. For
each lattice size the number of independent samples is 5000. The range
of the coupling constant $K$ in which we measure the observables depends on
the models, and they are shown in table \ref{tab:4.1}. 
%
\begin{table}[ht]
 \begin{center}
 \begin{tabular}{| c | c | c |}\hline
      \multicolumn{1}{|c|}{$~~Q~~$} & 
      \multicolumn{1}{c|}{$~~$ range of $K$ $~~$} & 
      \multicolumn{1}{c|}{$~~~~$ interval of $K$ $~~~~$} \\ 
 \hline
   0.2 & 0.25 $\sim$ 0.82 & 0.03 $\times$ 20 (points) \\
   0.4 & 0.50 $\sim$ 1.07 & 0.03 $\times$ 20 (points) \\
   0.6 & 0.70 $\sim$ 1.27 & 0.03 $\times$ 20 (points) \\
   0.8 & 0.80 $\sim$ 1.37 & 0.03 $\times$ 20 (points) \\
   1.0 & 0.83 $\sim$ 1.40 & 0.03 $\times$ 20 (points) \\
   1.5 & 1.05 $\sim$ 1.62 & 0.03 $\times$ 20 (points) \\
   2.0 & 1.23 $\sim$ 1.65 & 0.03 $\times$ 15 (points) \\
   2.5 & 1.30 $\sim$ 1.87 & 0.03 $\times$ 20 (points) \\
   3.0 & 1.42 $\sim$ 1.84 & 0.03 $\times$ 15 (points) \\
   3.5 & 1.45 $\sim$ 2.02 & 0.03 $\times$ 20 (points) \\
   4.0 & 1.55 $\sim$ 1.97 & 0.03 $\times$ 15 (points) \\
 \hline
 \end{tabular}
 \end{center}
 \caption{The ranges and intervals of the coupling constant $K$
          in which we measure the finite-size percolation probability 
          $P_N (K)$ for various $Q$.}
 \label{tab:4.1}
\end{table}
So far as the behavior of $P_N (K)$ is concerned, the order of phase 
transition is consistent with second order.

\begin{figure}[t]
 \vspace*{-0.5cm}
 \begin{tabular}{cc}
 \begin{minipage}{7cm}
   \epsfxsize=7cm \epsfbox {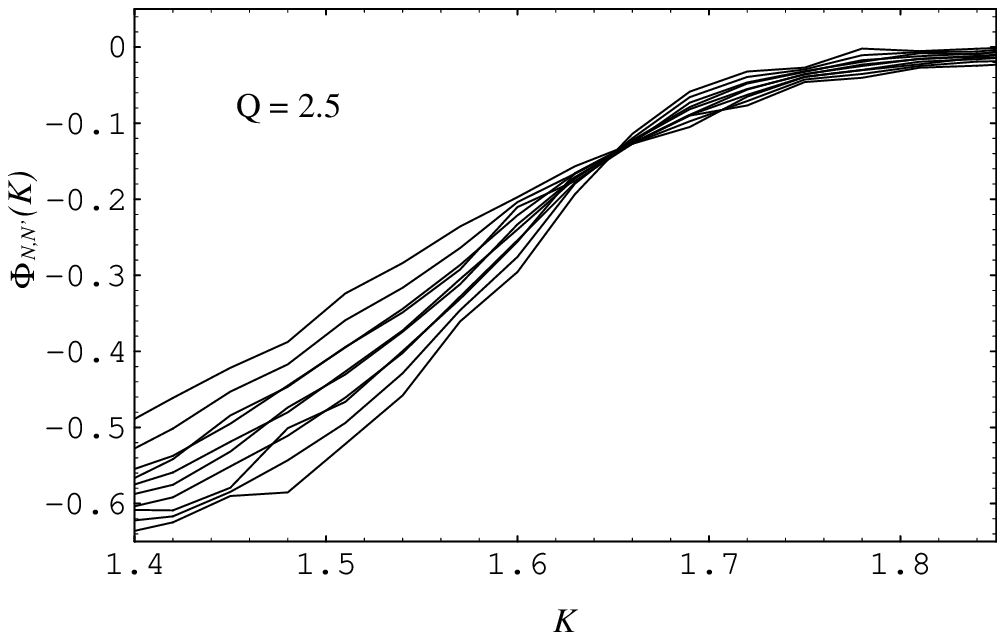}
 \end{minipage} &
 \begin{minipage}{7cm}
   \epsfxsize=7cm \epsfbox {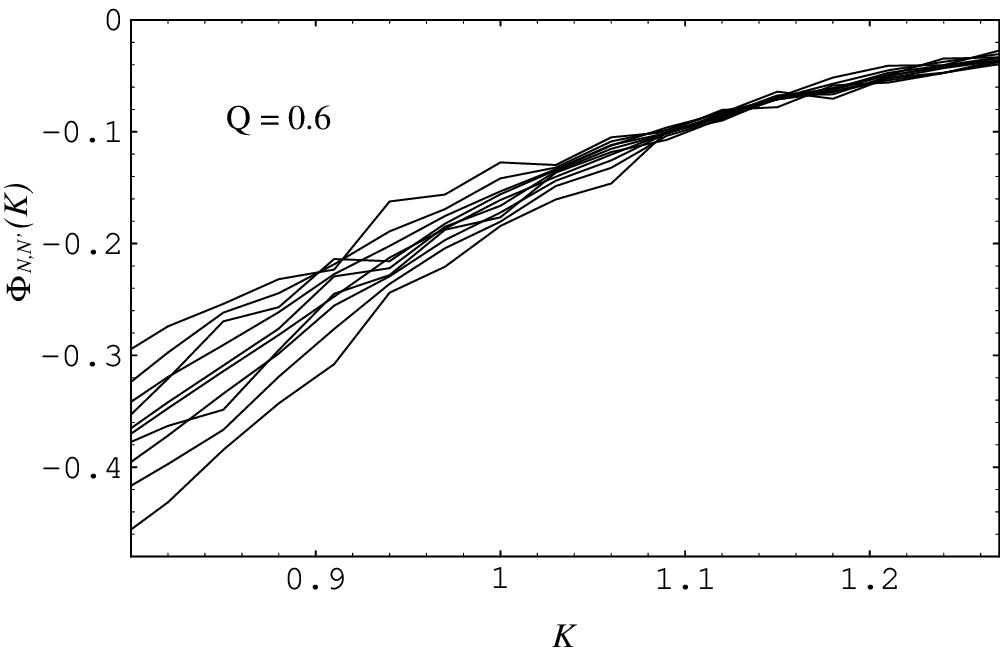}
 \end{minipage} 
 \end{tabular}
 \vspace*{-1.0cm}
 \caption{The functions $\Phi_{N,N'} (K)$, defined by eq.(\ref{eq:4.28}) 
          for all pairs of $N_i$'s where $N=100,\, 200, \cdots , 1600$.
          The curves are $Q=2.5$ and $0.6$ as examples.}
 \label{fig:4.4}
\end{figure}

For the behavior of $P_N (K)$ near $K\to K_c$ we suppose the following
scaling behavior based on the finite-size scaling hypothesis \cite{FB}
\begin{equation}
 \label{eq:4.27}
   P_N (K) =
   L^{-\beta_p /\nu}\, F_P [(K-K_c)\, L^{1 /\nu}]\, ,
\end{equation}
where (\ref{eq:4.26}) is assumed for the infinite system of $N\to \infty$.
$L$ is the 
linear extension of the system and $\nu$ is the critical exponent 
associated to the correlation length $\xi (K)\sim 1/|K-K_c|^{\nu}$.
Since the total volume is proportional to $N$, we should  
make an identification $N=L^{d_F}$ with $d_F$ as fractal dimension.
We may view $\xi (K)$ as an average geodesic distance. 
In order to extract $K_c$ using eq.(\ref{eq:4.27}), we define the
following function $\Phi_{N,N'} (K)$ \cite{BS}
\begin{equation}
 \label{eq:4.28}
   \Phi_{N,N'} (K) =
   \frac{\ln [P_N (K) /P_{N'} (K)]}{\ln [N / N']}\, ,
\end{equation}
for a pair of size $(N,N')$.
The functions
$\Phi_{N,N'} (K)$ and $\Phi_{N',N''} (K)$ for two different pairs of 
sizes $(N,N')$ and $(N',N'')$ should thus intersect at $K_c$,
and the intersection point should yield $-\beta_p / \nu d_F$ if corrections
to finite-size scaling can be neglected. 
In fig.\ref{fig:4.4} we plot
the functions $\Phi_{N,N'} (K)$ for all pairs of given sizes in the cases of
$Q=2.5$ and $0.6$. 
As we can see from fig.\ref{fig:4.2}, 
$P_N (K)$ grows with the increase of $Q$, while the intersection point
is relatively clear for larger $Q$. It is getting more 
difficult to find the intersection point for smaller $Q$.
%

Let us next define a quantity $n_s (K) = p_s (K)/s$, where $p_s (K)$ 
is a probability of a vertex being on a cluster of size $s$. 
$n_s (K)$ can be recognized as a cluster size distribution.
This can be understood as follows: 
Suppose we have $m_s$ clusters with seize $s$, we can obtain a relation 
$p_s (K)=m_s s/N$ and thus $n_s (K) = p_s (K)/s = m_s /N $ which is 
proportional to the cluster number and thus can be understood as a cluster 
size distribution. 
Using the cluster size distribution we define the so-called percolation 
susceptibility $\chi (K)$
\begin{equation}
 \label{eq:4.29}
   \chi (K) = \sum_s\! {}'\, s^2\, n_s (K)\, ,
\end{equation}
where the prime-sum means that the maximum cluster is omitted from the
summation. This quantity is the average number of bonds of a (finite)
cluster and is expected to show the following critical singularity in 
the infinite lattice for $|K-K_c|\to 0$  
\begin{equation}
 \label{eq:4.30}
   \chi (K) \sim \left\{
    \begin{array}{ll}
      (K-K_c)^{-\gamma_p}\, , & \quad K \geq K_c \\
      (K_c-K)^{-\gamma_p '}\, , & \quad K \leq K_c ~,
    \end{array} \right.
\end{equation}
where $\gamma_p$ and $\gamma_p '$ are the critical exponents associated 
to the percolation susceptibility.
Since these quantities $P(K)$ and $\chi (K)$ play a major role in usual 
percolation theory, we use them as crucial quantities of determining 
the critical point of generalized Potts models.

%
\begin{figure}[t]
 \vspace*{-0.5cm}
 \begin{tabular}{cc}
 \begin{minipage}{7cm}
   \epsfxsize=7cm \epsfbox {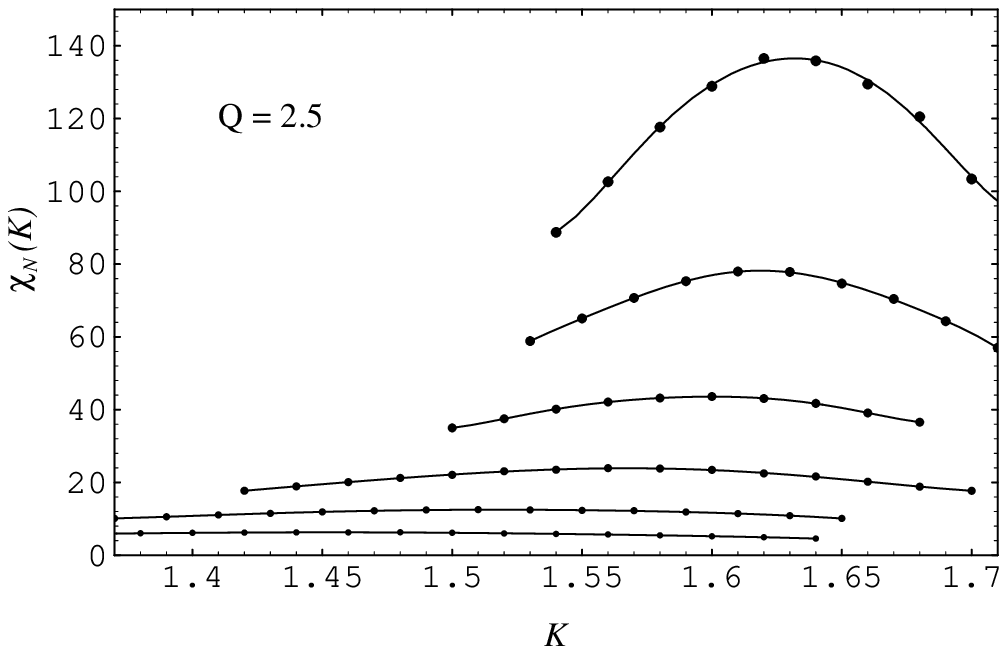}
 \end{minipage} &
 \begin{minipage}{7cm}
   \epsfxsize=7cm \epsfbox {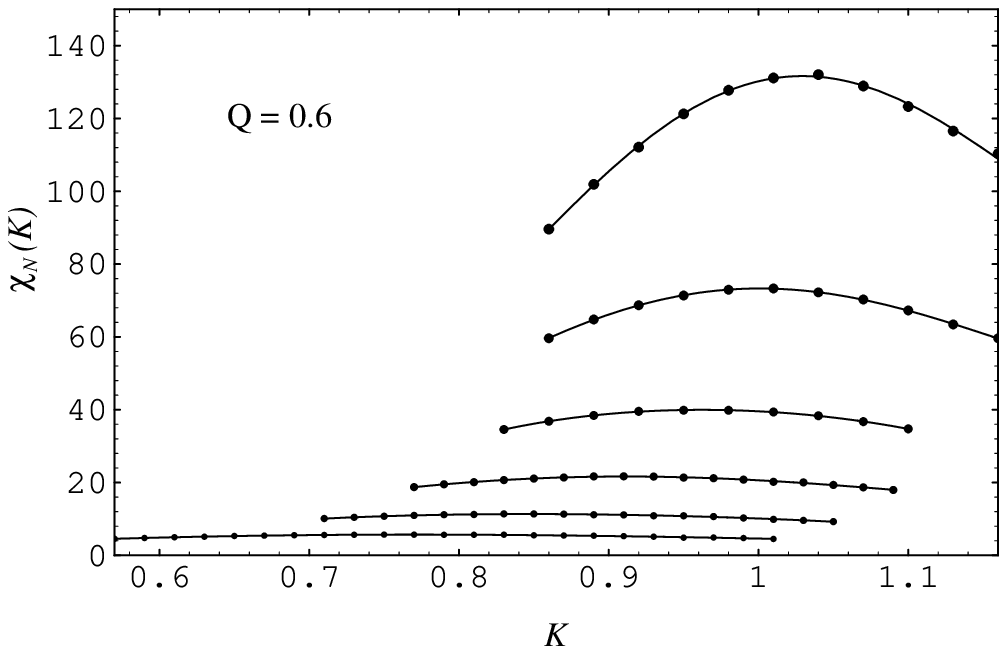}
 \end{minipage} 
 \end{tabular}
 \vspace*{-1.0cm}
 \caption{Size-dependence of $\chi_N(K)$ for $Q=2.5$ and $0.6$ as examples. 
          The sizes of system are $N=100$ (the lowest
          peak), $200,\, 400,\, 800,\, 1600$ and $3200$ 
          (the highest peak).}
 \label{fig:4.6}
\end{figure}
%
%
\begin{figure}[t]
 \vspace*{-0.5cm}
 \begin{tabular}{cc}
 \begin{minipage}{7cm}
   \epsfxsize=7cm \epsfbox {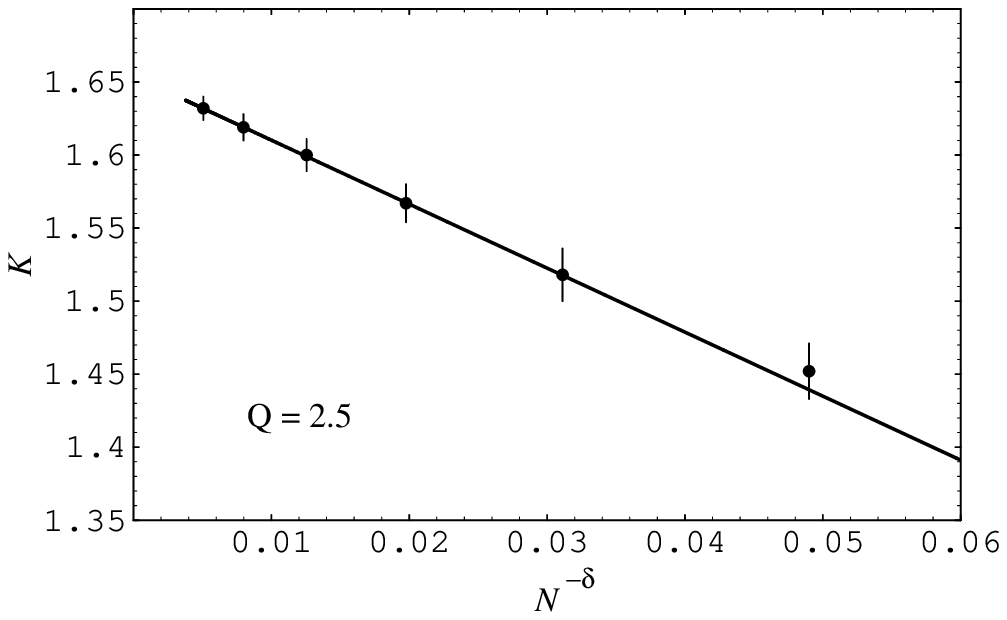}
 \end{minipage} &
 \begin{minipage}{7cm}
   \epsfxsize=7cm \epsfbox {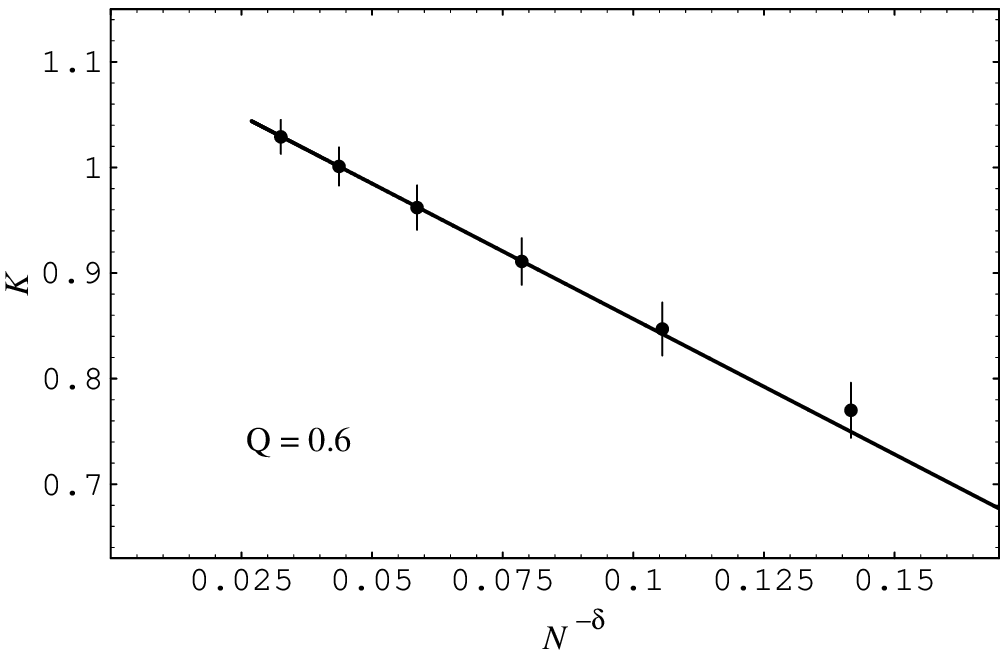}
 \end{minipage} 
 \end{tabular}
 \vspace*{-1.0cm}
 \caption{The best linear fit to eq.(\ref{eq:4.32}) for 
          $Q=2.5$ and $0.6$, where $\delta = 1/\nu d_F$.}
 \label{fig:4.8}
\end{figure}

In a finite lattice with the size $N$, 
the finite-size percolation susceptibility $\chi_N (K)$ can not diverge 
at the critical point $K_c$ 
but reaches a maximum of finite height only. The magnitude of this
maximum depends on the size of the lattice. 
In fig.\ref{fig:4.6} we show
the size dependence of $\chi_N (K)$ for $Q=2.5$ and $0.6$ as examples. 
In these simulations we take lattice sizes 
$N=100,\, 200,\, 400,\, 800,\, 1600$ and $3200$ for various values of $Q$'s. 
The range of the coupling constant $K$ and the number of independent 
samples in 
which we measure the observables depend on
the models and the lattice sizes, and they are shown in tables \ref{tab:4.2},
\ref{tab:4.3} and \ref{tab:4.4}.
In a finite lattice there is so called pseudo-critical point 
$\tilde K_c (N)$ instead of the true critical point $K_c$, where 
$\chi_N (K)$ reaches the maximum. The finite size scaling hypothesis for the
correlation length \cite{FB} means that at the pseudo-critical point 
the correlation length coincides with the linear extension of the 
system, i.e. 
\begin{equation}
 \label{eq:4.31}
   \xi (\tilde K_c (N)) \sim L \sim  N^{1/d_F}\, ,
\end{equation}
which, using $\xi (K)\sim |K-K_c|^{-\nu}$, yields for $N\to \infty$
\begin{equation}
 \label{eq:4.32}
   | \tilde K_c (N) - K_c | \sim N^{-1/\nu d_F}\, .
\end{equation}
In order to extract $K_c$ we make three-parameter fit to 
eq.(\ref{eq:4.32}). 
In fig.\ref{fig:4.8} we show the best linear fits 
for $Q=2.5$ and $0.6$. 
The error of $\tilde K_c (N)$ for each lattice size have 
been estimated using a polynomial fit to $\tilde K_c (N)$. Extrapolating to
$N=\infty$ yields the true critical coupling $K_c$.
%
\begin{figure}[t]
 \vspace*{-2cm}
 \begin{center}
 \begin{minipage}{12cm}
   \epsfxsize=12cm \epsfbox {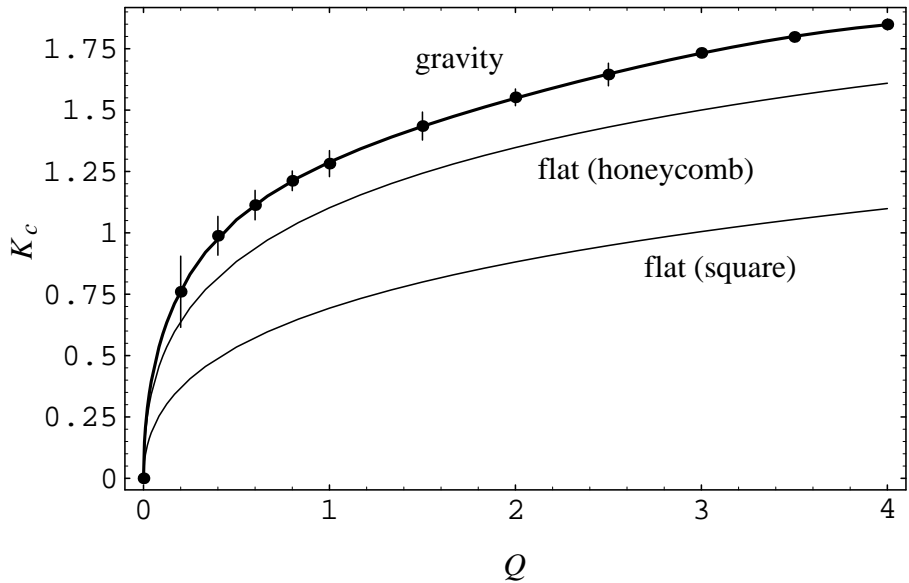}
 \end{minipage}
 \end{center}
 \vspace*{-2.5cm}
 \caption{The best values of $K_c$ versus $Q$.
          The fat solid line is the fitting curve to the data points with 
          polynomials of fifth-order in $Q^{1/2}$ (gravity). We also show the 
          theoretical critical 
          coupling for the $Q$-state Potts models on the honeycomb 
          and square lattice (flat).}
 \label{fig:4.10}
\end{figure}

For large $Q$ the critical couplings $K_c$ determined by using two methods 
as above are in agreement within the error. The discrepancy 
in the values is presumably due to the fact that larger lattices are
needed before the methods converge completely. In fig.\ref{fig:4.10} we 
have plotted the best values of $K_c$ versus values of $Q$'s. 
In this figure we have drawn the fitting function to data points with a
polynomial of fifth-order in $Q^{1/2}$.
We also show the theoretical critical couplings for 
the $Q$-state Potts models on the honeycomb \cite{KMS,KJ} and 
square \cite{P} lattice 
(i.e. no longer coupled to gravity). 
As these two curves show a similar trend, this figure 
suggests the existence of exact solutions for the $Q$-state Potts models 
coupled to gravity.

In the following subsections we use these values of the critical value $K_c$ 
obtained from the best-fit curve of fig.\ref{fig:4.10}.

\subsection{The string susceptibility }
The string susceptibility exponent $\gamma_s$ is one of the simplest 
quantities which characterize the fractal structure of quantum gravity.
This quantity is introduced as the exponent of the subleading
correction to the canonical partition function for random surfaces of fixed 
volume $A$
\begin{equation}
 \label{eq:4.33}
   Z(A) \sim e^{\Lambda_c A}\, A^{\gamma_s -3}\, ,
\end{equation}
for $A\to \infty$, where $\Lambda_c$ denotes the critical cosmological 
constant. 
As we show in eq.(\ref{eq:2.52-1}), the total area $A$ is proportional 
to the number of triangles $N$ in the regularized counterpart of the 
eq. (\ref{eq:4.33}). 
In the case of pure gravity, $Z(A=N)$ ($a^2=1$) is equal to 
the number of inequivalent triangulations with volume $N$.

Physically the string susceptibility can be identified as an 
order parameter for the branching probability of random surface. 
This could be understood by the following relation:
\begin{equation}
 \label{eq:4.33-1}
\frac{1}{Z(A)}\int_0^AdB BZ(B)(A-B)Z(A-B)\, \sim \, A^{\gamma_s}, 
\end{equation}
where $BZ(B)$ can be identified as a possible number of triangulation 
with a marked point on a triangulated surface of area $B$. 
Thus the left hand side of eq. (\ref{eq:4.33-1}) measures the average 
branching rate of the total surface branching into two parts\cite{Kaw}. 
For $\gamma_s>0$ the surface has tendency to branch more while 
for $\gamma_s<0$ the surface tends to be smooth. 

Using the analytical approach in a continuum framework, 
$c$-dependence of $\gamma_s$ was first derived by KPZ \cite{KPZ} 
and later rederived 
by DDK \cite{DDK} by using conformal gauge formulation of Liouville theory, 
\begin{equation}
 \label{eq:4.34}
   \gamma_s (c) =
   \frac{c-1-\sqrt{(25-c)(1-c)}}{12}\, ,
\end{equation}
for two-dimensional quantum gravity coupled to the matter central
charge $c$ with spherical topology. 
\begin{table}[t]
 \begin{center}
 \begin{tabular}{| c | c |}\hline
      \multicolumn{1}{|c|}{$~~~Q~~~$} & 
      \multicolumn{1}{c|}{$~~~~$ $K_c$ $~~~~$} \\ 
 \hline
   0.00 & 0.000 \\
   0.05 & 0.428 \\
   0.20 & 0.764 \\
   0.50 & 1.053 \\
   0.80 & 1.212 \\
   1.00 & 1.288 \\
   1.50 & 1.434 \\
   2.00 & 1.547 \\
   2.50 & 1.647 \\
   3.00 & 1.732 \\
   3.50 & 1.800 \\
   4.00 & 1.848 \\
 \hline
 \end{tabular}
 \end{center}
 \caption{The critical coupling constants $K_c$ for various values of $Q$.}
 \label{tab:4.5}
\end{table}

In this subsection we investigate the string susceptibility exponent 
$\gamma_s$ by measuring the distributions of so-called baby universes
\cite{JM,AJT,AT}. 
It has already been pointed out that the numerical 
values of string susceptibility are in perfect agreement 
with the theoretical results (\ref{eq:4.34}) of $Q$-state Potts
models for integral values of $Q$'s. 
We then expect that the agreement will be perfect even for the non-integral 
values of $Q$'s. 
We intend to use the numerical investigations of $\gamma_s (c)$ as the 
cross check of the critical values of $K_c$ calculated in the previous 
subsection.

The branching probability of the surface with total area $A$ into 
$B$ and $A-B$ is given by the integrand of eq. (\ref{eq:4.33-1}). 
The lattice counterpart of this branching probability is given by 
\begin{eqnarray}
 \label{eq:4.36}
   b_N (B) & \sim & 
   \frac{3\, (B+1)\, Z(B+1)\, (N-B+1)\, Z(N-B+1)}{Z(N)} \nonumber \\
   & \sim & 
   N\, \biggl[B\, \biggl(1-\frac{B}{N}\biggr)\biggr]^{\gamma_s -2}\, 
   (1<<B<N),
\end{eqnarray}
where two baby universes are divided by single triangle in the current 
formulation of triangulations.  
We may call the smaller part of the minimum neck as a baby universe (minbu). 
In numerical simulations it is easy to find the shortest loops in a
given triangulation and then to enumerate the area of the corresponding
minbu's.

The simulations were done with lattice sizes $N=1000$ and $2000$ for various 
values of $Q$. For each lattice size the number of independent samples is 
$100K$. The values of $Q$'s and the critical couplings $K_c (Q)$ are
shown in table \ref{tab:4.5}.
%
\begin{figure}[t]
 \vspace*{-2cm}
 \begin{center}
 \begin{minipage}{12cm}
   \epsfxsize=12cm \epsfbox {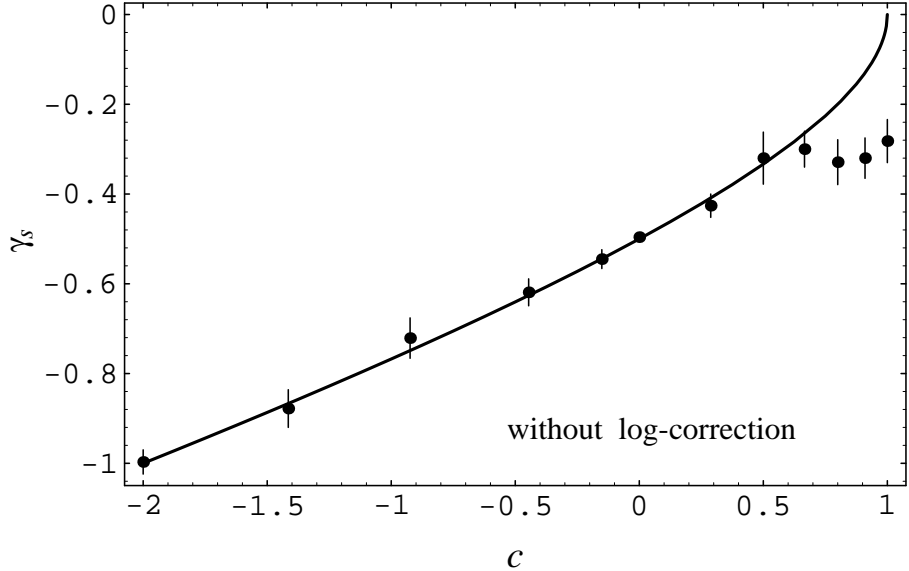}
 \end{minipage}
 \end{center}
 \vspace*{-2.5cm}
 \caption{The measured string susceptibility $\gamma_s$ versus central 
          charge $c$ and the theoretical curve. No logarithmic corrections 
          are introduced in the fits.}
 \label{fig:4.15}
\end{figure}

In order to extract $\gamma_s$ we have fitted the distributions 
expressed in the form
\begin{equation}
 \label{eq:4.37}
   \ln [ b_N (B) ] =
   a_1 + (\gamma_s -2)\, 
   \ln\biggl[B\, \biggl(1-\frac{B}{N}\biggr)\biggr] + \frac{a_2}{B}\, ,
\end{equation}
where $a_1$ and $a_2$ are fitting parameters and the last term is a 
finite size correction term for small $B$ \cite{AJT}. 
We have introduced a lower cut-off $B_c$, 
because eq.(\ref{eq:4.36}) is only asymptotically correct, deviations
can be expected for small $B$. Moreover we have cut large $B$ part, 
because the distributions of minbu's are not 
universal for large $B$. 

The values $\gamma_s (B_c)$ extracted from eq.(\ref{eq:4.37}) approach 
exponentially to a limiting value for large
$B_c$. Thus in order to extract the limiting value $\gamma_s$ we fit  
the values $\gamma_s (B_c)$ in such a form
\begin{equation}
 \label{eq:4.38}
   \gamma_s (B_c) = \gamma_s - a_1\, e^{-a_2 B_c}\, .
\end{equation}
In fig.\ref{fig:4.15} we plotted the limiting value $\gamma_s$ obtained by 
the above method versus central charge $c$ together with 
the theoretical curve (\ref{eq:4.34}).
%
\begin{figure}[t]
 \vspace*{-2cm}
 \begin{center}
 \begin{minipage}{12cm}
   \epsfxsize=12cm \epsfbox {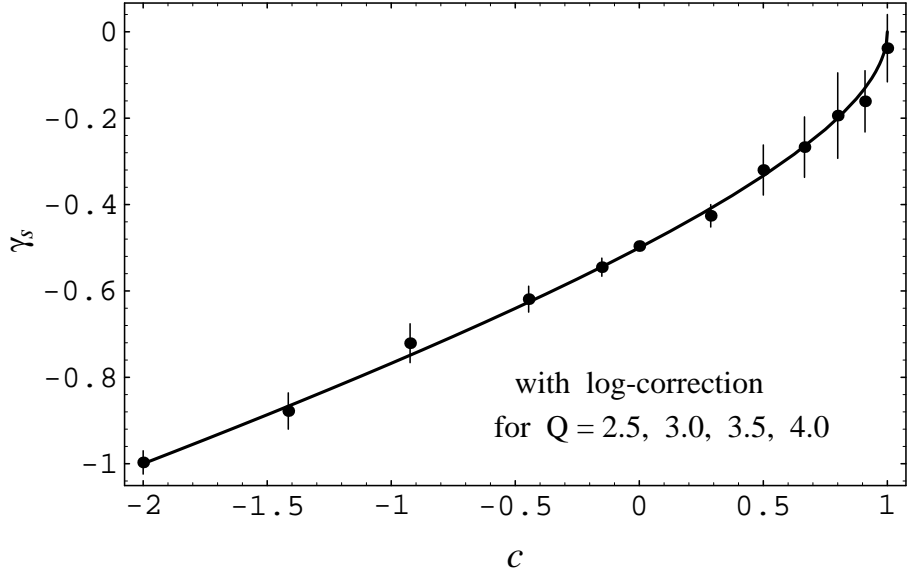}
 \end{minipage}
 \end{center}
 \vspace*{-2.5cm}
 \caption{The measured string susceptibility $\gamma_s$ versus central 
          charge $c$ and the theoretical curve. Logarithmic corrections 
          are introduced in the fits for $Q=2.5,\, 3.0,\, 3.5$ and $4.0$.}
 \label{fig:4.17}
\end{figure}

The reason why the results for $c\approx 1$ disagree with the theoretical
curve is possibly due to the fact that logarithmic corrections are not yet 
introduced. 
It is well known \cite{JM,AT} that logarithmic corrections play an 
important role in the vicinity of $c\approx 1$.
Thus we assume that the partition function has the following asymptotic
behavior for large $N$
\begin{equation}
 \label{eq:4.39}
   Z(N) \sim e^{\lambda_c N}\, N^{\gamma_s -3}\, (\ln N)^{\alpha}\, ,
\end{equation}
where $\alpha$ is an additional parameter.
The measured distributions $b_N (B)$ can now be fitted to the following 
parameterization:
\begin{eqnarray}
 \label{eq:4.40}
   \ln [ b_N (B) ] & = &
   a_1 + (\gamma_s -2)\, 
   \ln\biggl[B\, \biggl(1-\frac{B}{N}\biggr)\biggr] \nonumber\\ 
   && \mbox{}+ \alpha\, \ln [\ln B\cdot \ln (N-B)] + \frac{a_2}{B}\, ,
\end{eqnarray}
for $Q=2.5,\, 3.0,\, 3.5$ and $4.0$. 
Then $\gamma_s$ with the logarithmic corrections versus 
central charge $c$ are plotted in fig.\ref{fig:4.17}. 

The string susceptibility $\gamma_s$ with the logarithmic corrections 
for the various values of $Q$'s are in very good agreement with the 
theoretical curve. We can then conclude that the values of critical coupling 
$K_c (Q)$ evaluated numerically in the previous subsection are correct 
within errors.

\subsection{The fractal dimension}

The most straightforward definition of the fractal dimension is 
given by 
\begin{equation}
 \label{eq:2.7}
   N(r) = r^{d_F}, 
\end{equation}
where we count the number of triangles $N(r)$ inside 
$r$ steps from a marked triangle. 
In fact the fractal structure of the two-dimensional quantum gravity 
was first confirmed in this way by Kawamoto, Kazakov, Saeki and Watabiki 
for $c=-2$ scalar fermion model\cite{KKSW}. 
In these analyses they needed 5 million triangles to obtain the reliable 
value of the fractal dimension. 
It was later recognized that finite size scaling hypothesis is very useful 
to evaluate the fractal dimension numerically and thus relatively small 
number of triangles is enough to obtain very accurate fractal dimension 
for $c=-2$ model\cite{KWY1},\cite{KWY2}. 
Here we use finite size scaling hypothesis to obtain the $c$-dependence 
of the fractal dimension. 

It has already been recognized numerically in \cite{KKSW} that the 
total perimeter length $s_N(r)$ measured at geodesic distance $r$ from a 
marked triangle grows 
\begin{equation}
 \label{eq:2.8}
   s_N(r) = r^{d_F-1}. 
\end{equation}
This fact triggered the investigation of analytic derivation of transfer 
matrix for two dimensional random surface of quantum gravity for 
$c=0$ model\cite{KKMW}. 
It has then been recognized that two point function of random surface can be 
related to the measurement of $s_N(r)$ \cite{AW,AJW,ADJ}. 

In the case of pure gravity ($c=0$) "two-point function" with fixed volume 
$A$ is defined by
\begin{equation}
 \label{eq:2.11}
   S_A (R) = \frac{1}{Z(A)}\, \int {\cal D} [g] \, 
   \delta (\int d^2x \sqrt{g}-A)\,
   \frac{1}{A}\, \int \! d^2 x \sqrt{g} \,
   \int \! d^2 x' \sqrt{g} \, \delta (D_g (x,x') -R)\, ,
\end{equation}
where $D_g (x,x')$ denotes the geodesic distance between the 
points labeled by $x$ and $x'$ measured with respect to $g$.
Then $S_A (R)\, dR$ is the average area of a spherical shell of thickness
$dR$ and radius $R$ from a marked point on the manifold. 
We recognize that the lattice triangulation version of $S_A (R)$ 
corresponds to $s_N(r)$.
According to the numerical result, we expect to have a relation  
\begin{equation}
 \label{eq:2.13}
   S_A(R)  \sim  R^{d_F-1} ~~~~\mbox{for \ $R \sim 0$}\, .
\end{equation}

In case of pure gravity ($c=0$) $S_A(R)$ was calculated analytically 
\begin{equation}
 \label{eq:2.14}
   S_A(R) = R^3 f(R/A^{\frac{1}{4}})\, ,
\end{equation}
where $f(x) \sim e^{-x^{4/3}}$ for large $x$ \cite{AW,AJW}.
This analytic result is consistent with the calculation of the 
fractal dimension $d_F=4$ for $c=0$ model derived by transfer matrix 
formalism \cite{KKMW}. 
This result, however, strongly suggests the following scaling hypothesis 
for general matter central charge $c$ coupled two dimensional quantum 
gravity: 
\begin{equation}
 \label{eq:2.15}
   S_A(R) = A^{1-1/d_F} F(R/A^{1/d_F})\, ,
\end{equation}
where 
\begin{equation}
 \label{eq:2.16}
   F(x) \sim x^{d_F-1},~~~~x \ll 1\, .
\end{equation}
This beautiful scaling behavior was observed with a high accuracy 
in the very systematic numerical analyses of the fractal structure of 
two-dimensional quantum 
gravity coupled to $c=-2$ matter\cite{KWY1,KWY2}. 
Here we assume that the scaling hypothesis works in general for 
the general matter central charge $-2\leq c\leq 1$. 

%
%
\begin{figure}[t]
 \vspace*{-0.5cm}
 \begin{tabular}{cc}
 \begin{minipage}{7cm}
   \epsfxsize=7cm \epsfbox {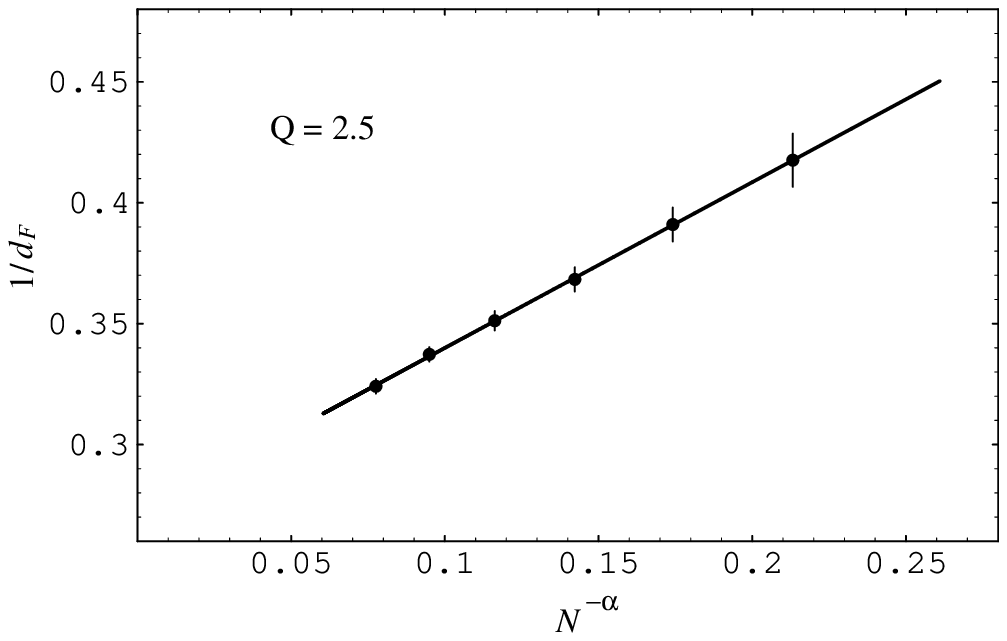}
 \end{minipage} &
 \begin{minipage}{7cm}
   \epsfxsize=7cm \epsfbox {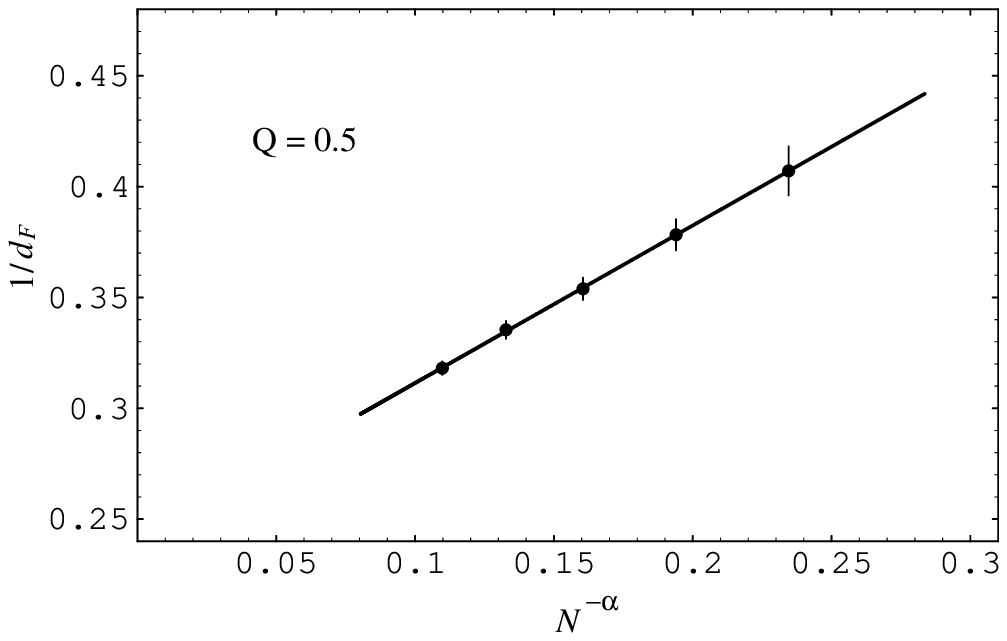}
 \end{minipage} 
 \end{tabular}
 \vspace*{-1.0cm}
 \caption{The best linear fit to eq.(\ref{eq:4.44}) for 
          $Q=2.5$ and $0.5$ as examples, 
          by the decay of the peak of $s_N(r)/N$ with $N$.}
 \label{fig:4.20}
\end{figure}
%
%
\begin{figure}[h]
 \vspace*{-0.5cm}
 \begin{tabular}{cc}
 \begin{minipage}{7cm}
   \epsfxsize=7cm \epsfbox {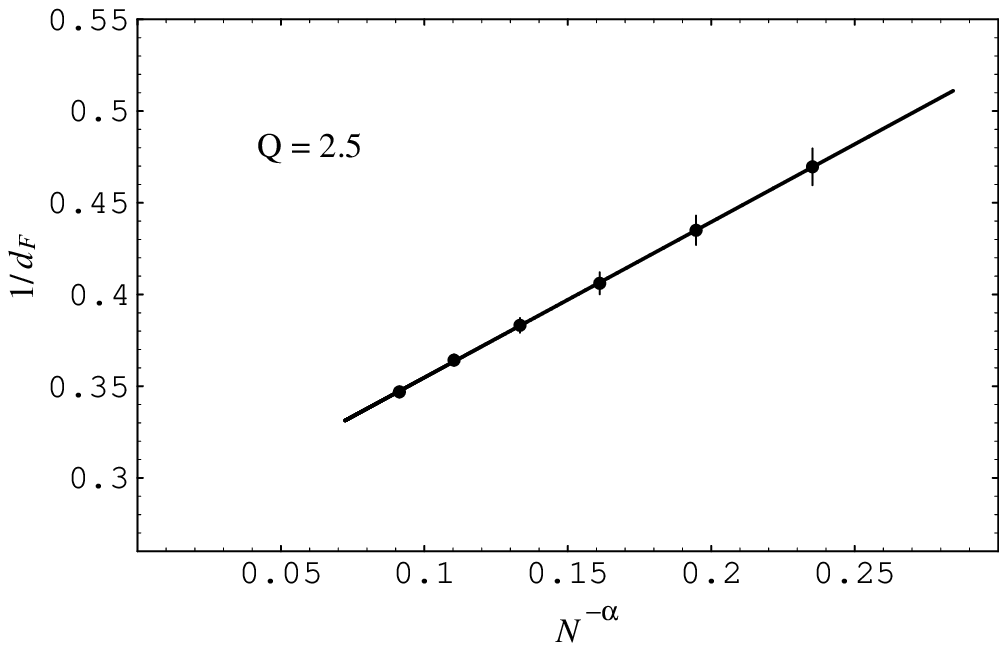}
 \end{minipage} &
 \begin{minipage}{7cm}
   \epsfxsize=7cm \epsfbox {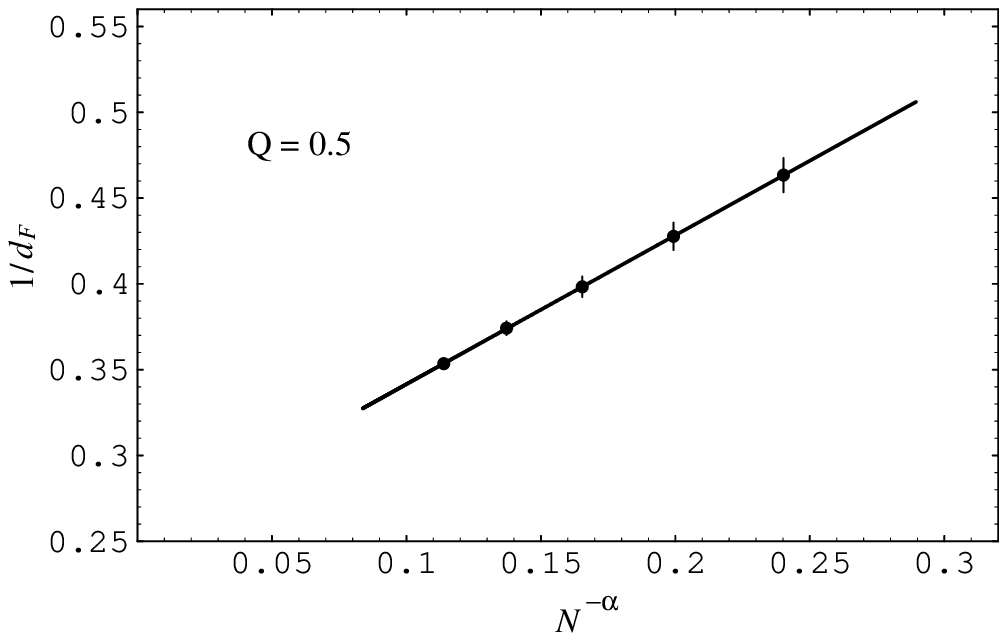}
 \end{minipage} 
 \end{tabular}
 \vspace*{-1.0cm}
 \caption{The best linear fit to eq.(\ref{eq:4.45}) for 
          $Q=2.5$ and $0.5$ as examples,
          by the decay of the inverse average radius 
          $1/\langle r\rangle_N$ with $N$.}
 \label{fig:4.23}
\end{figure}

In our numerical simulations we define a geodesic distance as a 
``triangle distance'' instead of link distance on a triangulation for 
saving the CPU time.
The triangle distance denotes the shortest
path along neighboring triangles between the two separate triangles. 
Therefore the triangle distance is equal to the ``edge distance'' in the dual
$\varphi^3$-graph. 
Then the discretized analogue of $S_A(R)$, $s_N (r)$ with 
$A=Na^2 (a^2=1)$, consists 
of all triangles with triangle distance $r$ measured from a marked 
triangle. Corresponding to the above continuum description
we expect the following behavior for $s_N(r)$:
\begin{equation}
 \label{eq:4.41}
   s_N (r) \sim N^{1-1/d_F}\, F(x)\, ,~~~~~x = \frac{r}{N^{1/d_F}}\, , 
\end{equation}
and we expect $F(x)$ to behave as $x^{d_F-1}$ for small $x$.

In order to measure correlation function $s_N (r)$ the simulations were 
carried out with lattice sizes ranging; 
$N=100,\, 200,\, 400,\, 800,\, 1600,\,
3200$ and $6400$ for $Q=0.0,\, 0.05,\, 0.2,\, 0.5$ and $0.8$.
For $Q=1.0,\, 1.5,\, 2.0,\, 2.5,\, 3.0,\, 3.5$ and $4.0$ we have 
performed the simulations with lattice sizes ranging; $N=100\sim 6400$
and $12800$.
For each lattice size ranging; $N=100\sim 1600$ the number of 
independent samples are $100K$, and we choose $20$ initial random 
vertices (rv) on each configuration. Then the number
of independent samples is $50K\times 40({\rm rv})$ for $N=3200$, 
$20K\times 100({\rm rv})$ for $N=6400$ and $10K\times 200({\rm rv})$ 
for $N=12800$, respectively.
The values of $Q$'s and its critical couplings $K_c (Q)$ are
shown in table \ref{tab:4.5}.
%
\begin{figure}[t]
 \vspace*{-2cm}
 \begin{center}
 \begin{minipage}{12cm}
   \epsfxsize=12cm \epsfbox {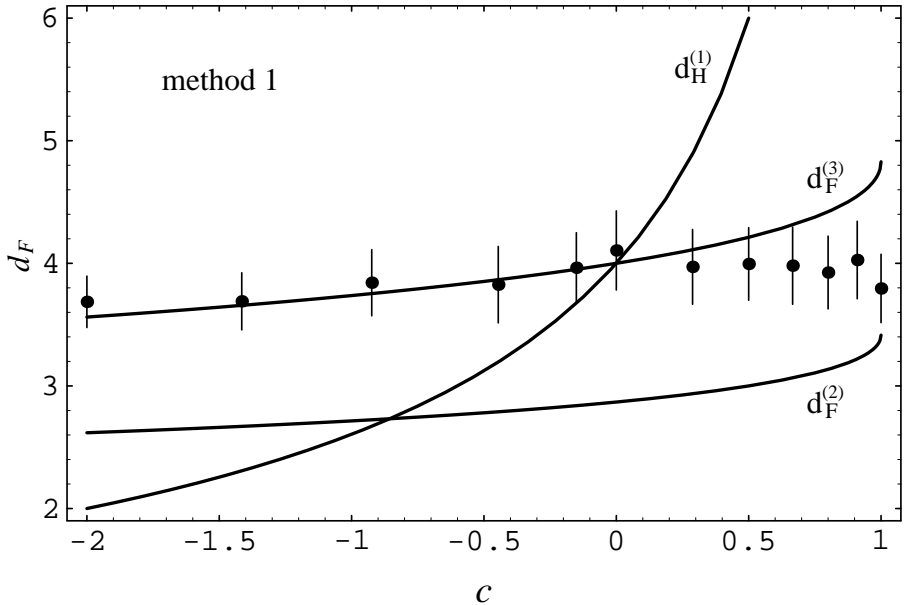}
 \end{minipage}
 \end{center}
 \vspace*{-2.5cm}
 \caption{The measured fractal dimension $d_F$ by the decay of peak 
          versus central charge $c$ and the three theoretical curves 
          given by eqs. (\ref{eq:2.60}), (\ref{eq:2.68}) and (\ref{eq:2.80}).}
 \label{fig:4.22}
\end{figure}
%
%
\begin{figure}[h]
 \vspace*{-2.0cm}
 \begin{center}
 \begin{minipage}{12cm}
   \epsfxsize=12cm \epsfbox {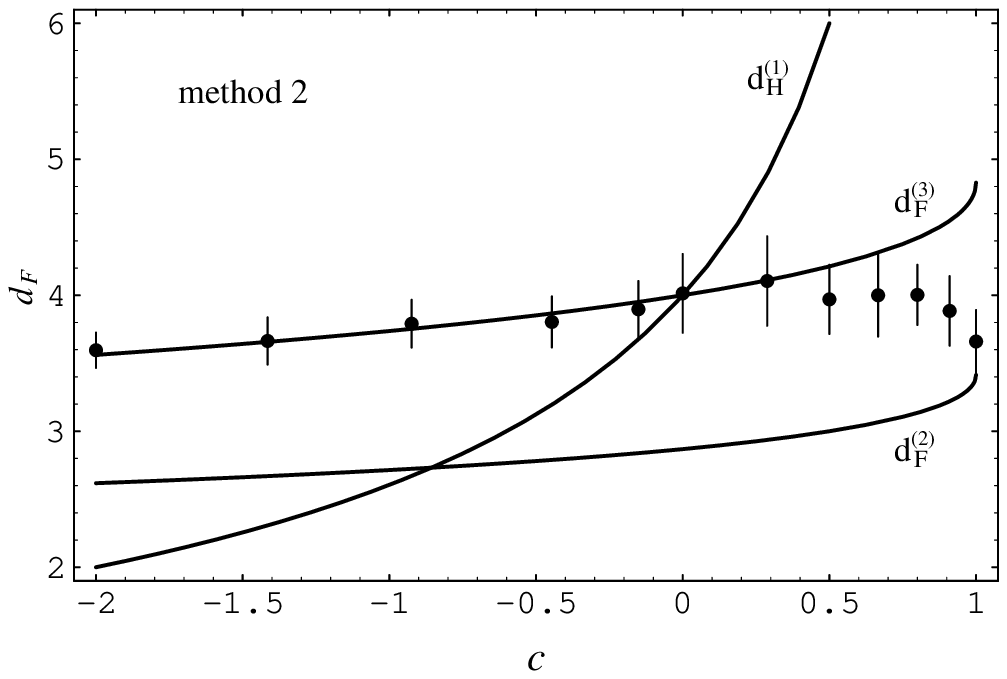}
 \end{minipage}
 \end{center}
 \vspace*{-2.5cm}
 \caption{The measured fractal dimension $d_F$ by the average radius 
          versus central charge $c$ and the three theoretical curves
          given by eqs. (\ref{eq:2.60}), (\ref{eq:2.68}) and (\ref{eq:2.80}).}
 \label{fig:4.25}
\end{figure}

In order to extract the fractal 
dimensions $d_F$ from the scaling relation (\ref{eq:4.41}),
It is crucial to introduce the so-called shift parameter 
$a$ \cite{AJW,KWY1,KWY2} which accommodate the finite size effects. 
This parameter 
$a$ is considered as the leading order correction in the scaling variable
$x=r / N^{1/d_F}$. This is reasonable from the point of view that
the shortest distances may include lattice artifacts, and thus  
we can not expect an exact
agreement with the continuum formulae. On the other hand, in order to 
incorporate the higher order corrections in the scaling variable $x$
we need to introduce a second shift parameter $b$ \cite{AJW}: 
$N^{1/d_F}\to N^{1/d_F} +b$. In this way we are led to a 
``phenomenological'' scaling variable $x$
\begin{equation}
 \label{eq:4.42}
   x = \frac{r+a}{N^{1/d_F} +b}\, . 
\end{equation}
In the present simulations with small lattice sizes  
(at most $N\sim 10^3$) we need to introduce the parameter $b$, however, 
it is numerically very hard to determine the optimal values of three 
parameters $a,\,b$ and $d_F$ in the case of the gravity coupled to matter. 
Although such optimal values of parameters are obtained, they keep the 
dependence on lattice size.

We have extracted the fractal dimensions $d_F$ by the following method.
First we use the decay of the peak of $s_N(r)/N$ with $N$ (this is similar 
to the use of finite-size scaling in the case of the percolation 
susceptibility). From eq.(\ref{eq:4.41}) we expect the 
following scaling behavior:
\begin{equation}
 \label{eq:4.43}
   s_N (r)/N \sim N^{-1/d_F}\, .
\end{equation}
We have divided $N$'s into three successive divisions such as 
$N$'s ($(100,200,400), \\
(200,400,800),$ etc.), and we have performed to fit 
$s_N (r)/N = a_1\, N^{-1/d_F}$ for each successive divisions by two 
parameters. In this way the $d_F$ dependence on $N$ becomes clear. 
Then we further assume the following scaling behavior:  
\begin{equation}
 \label{eq:4.44}
   |1/d_F(N) - 1/d_F(\infty)| \sim N^{-\alpha}\, ,
\end{equation}
where we can extrapolate $1/d_F(\infty)$ by a linear fit to 
eq.(\ref{eq:4.44}).
This is shown in fig.\ref{fig:4.20} for $Q=2.5$ and $0.5$ as examples, 
where the estimation of error is based on a non-linear fits to $s_N(r)/N$.
The values of $d_F$'s obtained in this way for the various values of $Q$'s
are plotted in fig.\ref{fig:4.22}. 
Three theoretical curves given by the formulae; (\ref{eq:2.60}), 
(\ref{eq:2.68}) and (\ref{eq:2.80}), are shown to be compared.
It is clear that the formula (\ref{eq:2.80}) is closer to the numerical 
values of the fractal dimension. 

Secondly, in order to make use of the whole information of $s_N(r)$
we have used the average radius $\langle r \rangle_N$ of universes with 
volume $N$
\begin{equation}
 \label{eq:4.45}
   \langle r \rangle_N \equiv \frac{1}{N}\sum_{r=0}^\infty\, r \; s_N(r)  
   \sim  N^{1/d_F}\, .
\end{equation}
We can then expect $1/\langle r \rangle_N \sim N^{-1/d_F}$, where $d_F$ 
can be determined in the same way as above. 
In fig.\ref{fig:4.23} we show the
corresponding linear fits to eq.(\ref{eq:4.45}). 
The values of $d_F$'s obtained in this way for the various values 
of $Q$'s are plotted in fig.\ref{fig:4.25}. 

These two independent results are consistent with each other and 
support the theoretical prediction $d_F^{(3)} (c)$.

\section{Conclusion and Discussions}
In this article we have shown the results of numerical investigations of 
the fractal dimension of two-dimensional quantum gravity coupled to 
the matter central charge $c$ for $-2\leq c\leq 1$. 
The $c$-dependence of the matter central charge is introduced by 
reformulating $Q$-state Potts model into the model which can be 
identified as a generalization of percolation cluster model, 
weighted percolation cluster model, on the random lattice.
In this formulation $Q$ can be generalized into non-integer value 
and thus continuous $c$-dependence is realized and then we have called this 
model simply as the generalized Potts model. 
Since the model has a percolation cluster feature, we have formulated a new 
Metropolis algorithm to generate clusters on dynamically triangulated 
surface.  

The $c$-dependence of the critical coupling $K_c$ is not known theoretically. 
We have evaluated the $c$-dependence of $K_c$ by measuring percolation 
probability and percolation susceptibility with the help of scaling 
hypothesis. The $c$-dependence of the critical coupling has similar behaviour 
as those of flat lattice. It is then very natural to expect that 
there is a theoretical solution of theoretical coupling $K_c$ in 
two-dimensional quantum gravity coupled to matter central charge $c$. 
The order of the phase transition is assumed to be second or even 
higher order at the critical point and we have not observed any evidence 
against this assumption. 
The string susceptibility is measured by the baby universe technique 
and has excellent agreement with the theoretical curve. It is recognized 
that the next leading logarithmic correction is important to improve the 
agreement with the theoretical prediction in particular for the region near 
$c\approx 1$.
Several parameters were needed to incorporate the finite size effects. 
The string susceptibility measurement is carried out as a cross check of 
the critical values of $K_c$ and the result was perfectly consistent with 
the independent measurement of the critical exponent in the above. 

The $c$-dependence of the fractal dimension is measured based on the 
two-point function of quantum surface with the finite size scaling 
hypothesis. We needed two extra parameters to accommodate the finite size 
corrections to the scaling parameter. 
Measurements are carried out by two methods; the decay behaviour of 
the peak of two-point function and the average radius of universes. 
The results agree well each other. 
The $c$-dependence of the fractal dimension is excellent agreement 
with the following theoretical prediction except for the region 
$c\approx 1$: 
\begin{equation}
\label{eq:6.1}
 d_F^{(3)}(c)  
 \ = \ 
 2 \times \frac{ \sqrt{ 25 - c } \, + \, \sqrt{ 49 - c }}
               {\sqrt{ 25 - c } \, + \, \sqrt{  1 - c }         }.   
\end{equation}
We consider that the deviation of the agreement of the numerical values of 
the fractal dimensions near $c\approx 1$ from the theoretical values is 
possibly due to the fact that the size of the lattice is 
not large enough to observe the possible discrete jump from the 
fractal dimension $c<1$  of eq.(\ref{eq:6.1}) to the value of 
branched polymer phase $d_F=2, \gamma_{s}=1/2$\cite{ADJ2,Dur,JK}. 

It is interesting to measure the change of fractal dimension very 
accurately in this delicate region near $c\approx 1$. 
The theoretical curve of the $c$-dependence of the fractal dimension 
has infinite slope at $c=1$ and then turns into imaginary value. 
We conjecture that the fractal phase of two dimensional quantum gravity 
$c<1$ turns into branched polymer phase in $1<c$. 
We expect that there is a discrete jump of the fractal dimension at 
$c=1$. In order to measure this discrete jump numerically we may need 
huge number of triangles in the simulation. 

It is interesting to compare with the measurement of the string susceptibility 
in three-dimensional simplicial gravity. In three dimensions tetrahedron 
is the fundamental simplex which corresponds to the triangle of 
two-dimensional quantum gravity. In three-dimensional dynamical 
triangulation the gravitational constant can be a free parameter and 
plays a role of central charge $c$ of two-dimensional quantum gravity. 
It was measured that the string susceptibility changes from negative 
region to the positive region where branched polymer phase is 
expected\cite{ABKW}. Here again the phase change from the fractal phase to 
the branched polymer phase is expected. 
For realistic higher dimensional simplicial quantum gravity it would be 
important to understand the phase change such as the fractal-branched polymer 
phase change.

\begin{table}[ht]
 \begin{center}
 \begin{tabular}{| c | c | c | c | c |}\hline
      \multicolumn{1}{|c|}{$~~Q~~$} & 
      \multicolumn{1}{c|}{\# triangles} & 
      \multicolumn{1}{c|}{$~~$ range of $K$ $~~$} & 
      \multicolumn{1}{c|}{$~~~~$ interval of $K$ $~~~~$} & 
      \multicolumn{1}{c|}{\# samples} \\
 \hline
   0.2 & 100 & 0.20 $\sim$ 0.58 & 0.02 $\times$ 20 (points) & 20000 \\
   {}  & 200 & 0.30 $\sim$ 0.58 & 0.02 $\times$ 15 (points) & 20000 \\
   {}  & 400 & 0.36 $\sim$ 0.64 & 0.02 $\times$ 15 (points) & 20000 \\
   {}  & 800 & 0.40 $\sim$ 0.67 & 0.03 $\times$ 10 (points) & 20000 \\
   {}  & 1600 & 0.48 $\sim$ 0.66 & 0.02 $\times$ 10 (points) & 20000 \\
   {}  & 3200 & 0.47 $\sim$ 0.74 & 0.03 $\times$ 10 (points) & 5000 \\
 \hline
   0.4 & 100 & 0.42 $\sim$ 0.80 & 0.02 $\times$ 20 (points) & 20000 \\
   {}  & 200 & 0.53 $\sim$ 0.81 & 0.02 $\times$ 15 (points) & 20000 \\
   {}  & 400 & 0.61 $\sim$ 0.89 & 0.02 $\times$ 15 (points) & 20000 \\
   {}  & 800 & 0.66 $\sim$ 0.93 & 0.03 $\times$ 10 (points) & 20000 \\
   {}  & 1600 & 0.75 $\sim$ 0.93 & 0.02 $\times$ 10 (points) & 20000 \\
   {}  & 3200 & 0.74 $\sim$ 1.01 & 0.03 $\times$ 10 (points) & 5000 \\
 \hline
   0.6 & 100 & 0.57 $\sim$ 1.01 & 0.02 $\times$ 23 (points) & 20000 \\
   {}  & 200 & 0.71 $\sim$ 1.05 & 0.02 $\times$ 18 (points) & 20000 \\
   {}  & 400 & 0.77 $\sim$ 1.09 & 0.02 $\times$ 17 (points) & 20000 \\
   {}  & 800 & 0.83 $\sim$ 1.10 & 0.03 $\times$ 10 (points) & 20000 \\
   {}  & 1600 & 0.86 $\sim$ 1.16 & 0.03 $\times$ 11 (points) & 20000 \\
   {}  & 3200 & 0.86 $\sim$ 1.19 & 0.03 $\times$ 12 (points) & 10000 \\
 \hline
   0.8 & 100 & 0.70 $\sim$ 1.14 & 0.02 $\times$ 23 (points) & 20000 \\
   {}  & 200 & 0.83 $\sim$ 1.11 & 0.02 $\times$ 15 (points) & 20000 \\
   {}  & 400 & 0.91 $\sim$ 1.19 & 0.02 $\times$ 15 (points) & 20000 \\
   {}  & 800 & 0.95 $\sim$ 1.22 & 0.03 $\times$ 10 (points) & 20000 \\
   {}  & 1600 & 0.97 $\sim$ 1.27 & 0.03 $\times$ 11 (points) & 20000 \\
   {}  & 3200 & 1.02 $\sim$ 1.29 & 0.03 $\times$ 10 (points) & 10000 \\
 \hline
 \end{tabular}
 \end{center}
 \caption{The lattice sizes, ranges of the coupling constant $K$
          and the number of samples
          in which we measure the finite-size percolation susceptibility
          $\chi_N (K)$ for $Q = 0.2,\, 0.4,\, 0.6$ and $0.8$.}
 \label{tab:4.2}
\end{table}
%
%
\begin{table}[ht]
 \begin{center}
 \begin{tabular}{| c | c | c | c | c |}\hline
      \multicolumn{1}{|c|}{$~~Q~~$} & 
      \multicolumn{1}{c|}{\# triangles} & 
      \multicolumn{1}{c|}{$~~$ range of $K$ $~~$} & 
      \multicolumn{1}{c|}{$~~~~$ interval of $K$ $~~~~$} & 
      \multicolumn{1}{c|}{\# samples} \\
 \hline
   1.0 & 100 & 0.82 $\sim$ 1.20 & 0.02 $\times$ 20 (points) & 20000 \\
   {}  & 200 & 0.94 $\sim$ 1.22 & 0.02 $\times$ 15 (points) & 20000 \\
   {}  & 400 & 1.00 $\sim$ 1.32 & 0.02 $\times$ 17 (points) & 20000 \\
   {}  & 800 & 1.05 $\sim$ 1.32 & 0.03 $\times$ 10 (points) & 20000 \\
   {}  & 1600 & 1.13 $\sim$ 1.31 & 0.02 $\times$ 10 (points) & 20000 \\
   {}  & 3200 & 1.14 $\sim$ 1.34 & 0.02 $\times$ 11 (points) & 10000 \\
 \hline
   1.5 & 100 & 1.01 $\sim$ 1.39 & 0.02 $\times$ 20 (points) & 20000 \\
   {}  & 200 & 1.13 $\sim$ 1.41 & 0.02 $\times$ 15 (points) & 20000 \\
   {}  & 400 & 1.19 $\sim$ 1.47 & 0.02 $\times$ 15 (points) & 20000 \\
   {}  & 800 & 1.28 $\sim$ 1.46 & 0.02 $\times$ 10 (points) & 20000 \\
   {}  & 1600 & 1.31 $\sim$ 1.49 & 0.02 $\times$ 10 (points) & 20000 \\
   {}  & 3200 & 1.33 $\sim$ 1.51 & 0.02 $\times$ 10 (points) & 10000 \\
 \hline
   2.0 & 100 & 1.10 $\sim$ 1.60 & 0.02 $\times$ 26 (points) & 20000 \\
   {}  & 200 & 1.20 $\sim$ 1.60 & 0.02 $\times$ 21 (points) & 20000 \\
   {}  & 400 & 1.28 $\sim$ 1.64 & 0.02 $\times$ 19 (points) & 20000 \\
   {}  & 800 & 1.37 $\sim$ 1.61 & 0.02 $\times$ 13 (points) & 20000 \\
   {}  & 1600 & 1.41 $\sim$ 1.63 & 0.02 $\times$ 12 (points) & 20000 \\
   {}  & 3200 & 1.45 $\sim$ 1.63 & 0.02 $\times$ 10 (points) & 10000 \\
 \hline
   2.5 & 100 & 1.26 $\sim$ 1.64 & 0.02 $\times$ 20 (points) & 20000 \\
   {}  & 200 & 1.37 $\sim$ 1.65 & 0.02 $\times$ 15 (points) & 20000 \\
   {}  & 400 & 1.42 $\sim$ 1.70 & 0.02 $\times$ 15 (points) & 20000 \\
   {}  & 800 & 1.50 $\sim$ 1.68 & 0.02 $\times$ 10 (points) & 20000 \\
   {}  & 1600 & 1.53 $\sim$ 1.71 & 0.02 $\times$ 10 (points) & 20000 \\
   {}  & 3200 & 1.54 $\sim$ 1.72 & 0.02 $\times$ 10 (points) & 10000 \\
 \hline
 \end{tabular}
 \end{center}
 \caption{The same of table \ref{tab:4.2} for $Q = 1.0,\, 1.5,\, 2.0$ and 
 $2.5$.}
 \label{tab:4.3}
\end{table}
%
 
%
\begin{table}[ht]
 \begin{center}
 \begin{tabular}{| c | c | c | c | c |}\hline
      \multicolumn{1}{|c|}{$~~Q~~$} & 
      \multicolumn{1}{c|}{\# triangles} & 
      \multicolumn{1}{c|}{$~~$ range of $K$ $~~$} & 
      \multicolumn{1}{c|}{$~~~~$ interval of $K$ $~~~~$} & 
      \multicolumn{1}{c|}{\# samples} \\
 \hline
   3.0 & 100 & 1.35 $\sim$ 1.73 & 0.02 $\times$ 20 (points) & 20000 \\
   {}  & 200 & 1.46 $\sim$ 1.74 & 0.02 $\times$ 15 (points) & 20000 \\
   {}  & 400 & 1.50 $\sim$ 1.78 & 0.02 $\times$ 15 (points) & 20000 \\
   {}  & 800 & 1.59 $\sim$ 1.77 & 0.02 $\times$ 10 (points) & 20000 \\
   {}  & 1600 & 1.60 $\sim$ 1.80 & 0.02 $\times$ 11 (points) & 20000 \\
   {}  & 3200 & 1.62 $\sim$ 1.80 & 0.02 $\times$ 10 (points) & 10000 \\
 \hline
   3.5 & 100 & 1.40 $\sim$ 1.78 & 0.02 $\times$ 20 (points) & 20000 \\
   {}  & 200 & 1.53 $\sim$ 1.81 & 0.02 $\times$ 15 (points) & 20000 \\
   {}  & 400 & 1.58 $\sim$ 1.86 & 0.02 $\times$ 15 (points) & 20000 \\
   {}  & 800 & 1.66 $\sim$ 1.84 & 0.02 $\times$ 10 (points) & 20000 \\
   {}  & 1600 & 1.68 $\sim$ 1.86 & 0.02 $\times$ 10 (points) & 20000 \\
   {}  & 3200 & 1.69 $\sim$ 1.87 & 0.02 $\times$ 10 (points) & 10000 \\
 \hline
   4.0 & 100 & 1.42 $\sim$ 1.90 & 0.02 $\times$ 25 (points) & 20000 \\
   {}  & 200 & 1.56 $\sim$ 1.90 & 0.02 $\times$ 18 (points) & 20000 \\
   {}  & 400 & 1.65 $\sim$ 1.93 & 0.02 $\times$ 15 (points) & 20000 \\
   {}  & 800 & 1.73 $\sim$ 1.91 & 0.02 $\times$ 10 (points) & 20000 \\
   {}  & 1600 & 1.74 $\sim$ 1.92 & 0.02 $\times$ 10 (points) & 20000 \\
   {}  & 3200 & 1.75 $\sim$ 1.93 & 0.02 $\times$ 10 (points) & 10000 \\
 \hline
 \end{tabular}
 \end{center}
 \caption{The same of table \ref{tab:4.2} for $Q = 3.0,\, 3.5$ and $4.0$.}
 \label{tab:4.4}
\end{table}

\vspace{2cm}   
\noindent{\Large \textbf{Acknowledgments}}\\

We would like to thank I. Kostov and Y. Watabiki for useful discussions at the 
very early stage of this work. This work is supported in part by Japanese 
Ministry of Education, Science, Sports and Culture under the grant number 
13640250.


\begin{thebibliography}{xx}
\addcontentsline{toc}{chapter}{References}
%
\bibitem{KPZ} 
   V. Knizhnik, A.M. Polyakov and A. Zamolodchikov,
   Mod. Phys. Lett. {\bf A3} (1988) 819.
%
\bibitem{DHK}
   J. Distler, Z. Hlousek and H. Kawai,
   Mod. Phys. Lett. {\bf A5} (1990) 1093.
%
\bibitem{KN}   
   H. Kawai and M. Ninomiya,
   Nucl. Phys. {\bf B336} (1990) 115.
%
\bibitem{AM}   
   M. E. Agishtein and A. A. Migdal, 
   Int. J. Mod. Phs. {\bf C1} (1990) 165; Nucl. Phys. {\bf B350} (1991) 690.
%
\bibitem{KKSW}
   N. Kawamoto, V.A. Kazakov, Y. Saeki and Y. Watabiki,
   Phys. Rev. Lett. {\bf 68} (1992) 2113. 
%
\bibitem{KSW}
   N. Kawamoto, Y. Saeki and Y. Watabiki, unpublished; 
   Y. Watabiki, 
   Progress in Theoretical Physics, Suppl. No. {\bf 114} (1993) 1; 
   N. Kawamoto, 
   in Proceedings of Nishinomiya-Yukawa workshop, Nishinomiya 1992,  
   {\sl Quantum gravity\/},
   ed. K. Kikkawa and M. Ninomiya (World Scientific) p. 112; 
   In First Asia-Pacific Winter School for Theoretical Physics 1993,
   Proceedings, {\sl Current Topics in Theoretical Physics\/}, 
   ed. Y.M. Cho (World Scientific).       
%
\bibitem{KKMW} 
   H. Kawai, N. Kawamoto, T. Mogami and Y. Watabiki,
   Phys. Lett. {\bf B306} (1993) 19. 
%
\bibitem{W}
   Y. Watabiki,
   Nucl. Phys. {\bf B441} (1995) 119; Phys. Lett. {\bf B346} (1995) 46.
%
\bibitem{AW}
   J. Ambj\o rn and Y. Watabiki,
   Nucl. Phys. {\bf B445} (1995) 129. 
%
\bibitem{AKW}
   J. Ambj\o rn, C.F. Kristjansen and Y. Watabiki,
   Nucl. Phys. {\bf B504} (1997) 555. 
%
\bibitem{OTY}
   S. Oda, N. Tsuda and T. Yukawa, Prog. Theor. Phys. {\bf 99} (1998) 875.
%
\bibitem{JM}
   S. Jain, and S.D. Mathur,
   Phys. Lett. {\bf B286} (1992) 239.
%
\bibitem{AJT}
   J. Ambj\o rn, S. Jain, and G. Thorleifsson,
   Phys. Lett. {\bf B307} (1993) 34.
%
\bibitem{AJW} 
   J. Ambj\o rn, J. Jurkiewicz and Y. Watabiki,
   Nucl. Phys. {\bf B454} (1995) 313. 
%
\bibitem{KWY1} 
   J. Ambj\o rn, K.N. Anagnostopoulos, T. Ichihara, 
   L. Jensen, N. Kawamoto, Y. Watabiki and K. Yotsuji, 
   Phys. Lett. {\bf B397} (1997) 177.
%
\bibitem{KWY2} 
   J. Ambj\o rn, K.N. Anagnostopoulos, T. Ichihara, 
   L. Jensen, N. Kawamoto, Y. Watabiki and K. Yotsuji, 
   Nucl. Phys. {\bf B511} (1998) 673.
%
\bibitem{AA}
   J. Ambj\o rn and K.N. Anagnostopoulos, 
   Nucl. Phys. {\bf B497} (1997) 445.
%
\bibitem{AT}
   J. Ambj\o rn and G. Thorleifsson,
   Phys. Lett. {\bf B323} (1994) 7.
%
\bibitem{P}
   R.B. Potts, Proc. Camb. Phil. Soc. {\bf 48} (1952) 106. 
%
\bibitem{FK}
   C.M. Fortuin and P.W. Kasteleyn,
   J. Phys. Soc. Jpn. Suppl. {\bf 26} (1969) 11;
   Physica {\bf 57} (1972) 536.
%
\bibitem{JKPS} 
   J. Jurkiewicz, A. Krzywicki, B. Petersson and B. Soderberg, 
   Phys. Lett. {\bf B213} (1988) 511; 
   C.F. Baillie and D.A. Johnston, Phys. Lett. {\bf B286} (1992) 44; 
   S. Catterall, J. Kogut and R. Renken, Phys. Lett. {\bf B292} (1992) 277; 
   J. Ambj\o rn, B. Durhuus, T. Jonsson and G. Thorleifsson, 
   Nucl. Phys. {\bf B398} (1993) 568;
   J. Ambj\o rn, G. Thorleifsson and M. Wexler,
   Nucl. Phys. {\bf B439} (1995) 187.
%
\bibitem{ATW}
   J. Ambj\o rn, G. Thorleifsson and M. Wexler,
   Nucl. Phys. {\bf B439} (1995) 187. 
%
%
\bibitem{DDK}
   F. David,
   Mod. Phys. Lett. {\bf A3} (1988) 651, \\
   J. Distler and H. Kawai,
   Nucl. Phys. {\bf B321} (1989) 509.
%
\bibitem{epsilon-expansion}
   S. Weinberg, {\it in} General Relativity, an Einstein centenary 
   Survey, ed. S. W. Hawking and W. Israel (Cambridge University 
   Press, 1979) p. 790,\\
   R. Gastmans, R. Kallosh and C. Truffin, 
   Nucl. Phys. {\bf B133} (1978) 417, \\
   S.M. Christensen adn M.J. Duff,
   Phys. Lett. {\bf B79} (1978) 213. 
%
\bibitem{Isi}
   E. Ising, Z. Physik {\bf 31} (1925) 253.
%
\bibitem{BKW}
   R.J. Baxter, S.B. Kelland and F.Y. Wu, J. Phys. {\bf A9} (1976) 397. 
%
\bibitem{Tut}
   W.T. Tutte, J. Combinatorial Theory {\bf 2} (1967) 301.
%
\bibitem{Wu}
   F.Y. Wu, J. Phys. {\bf A10} (1977) L113. 
%
\bibitem{TL}
   J.N.V. Temperley and E.H. Lieb,
   Proc. R. Soc. London Ser. {\bf A322} (1971) 251.
%
\bibitem{Dot}
   Vl.S. Dotsenko, Nucl. Phys. {\bf B235} (1984) 54. 
%
\bibitem{DF}
   Vl.S. Dotsenko and V.A. Fateev, Nucl. Phys. {\bf B240} (1984) 312. 
%
\bibitem{Gla}
   R.J. Glauber, J. Math. Phys. {\bf 4} (1963) 294. 
%
\bibitem{Swe}
   M. Sweeny, Phys. Rev. {\bf B27} (1983) 4445. 
%
\bibitem{BKKM}
   V.A. Kazakov, I.K. Kostov and A.A. Migdal, 
   Phys. Lett. {\bf B157} (1985) 295; 
   D.V. Boulatov, V.A. Kazakov, I.K. Kostov and A.A. Migdal, 
   Nucl. Phys. {\bf B275} (1986) 641. 
%
\bibitem{Kaz}
   V.A. Kazakov,
   Phys. Lett. {\bf A119} (1986) 140.
%
\bibitem{BK}
   D.V. Boulatov and V.A. Kazakov, 
   Phys. Lett. {\bf B186} (1987) 379.
%
\bibitem{Sta}
   D. Stauffer,
   {\sl Introduction to Percolation Theory\/},
   (Taylor \& Francis, 1985).
%
\bibitem{FB}
   M.E. Fisher and M.N. Barber,
   Phys. Rev. Lett. {\bf 28} (1972) 1516.
%
\bibitem{BS}
   M.N. Barber and W. Selke, J. Phys. {\bf A15} (1982) L617.
%
\bibitem{KMS}
   T. Kihara, Y. Midzuno and T. Shizume,
   J. Phys. Soc. Japan {\bf 9} (1954) 681.
%
\bibitem{KJ}
   D. Kim and R. Joseph, J. Phys. {\bf C7} (1974) L167.
%
\bibitem{Kaw}
   H. Kawai, Nucl. Phys. {\bf B26}(Proc. Suppl.) (1992) 93.
%
\bibitem{ADJ}
   J. Ambj\o rn, B. Durhuus and T. Jonsson,
   {\sl Quantum Geometry\/},
   (Cambridge University Press, 1997).
%
\bibitem{ADJ2}
   J. Ambj\o rn, B. Durhuus and T. Jonsson, 
   Phys. Lett. {\bf B244} (1990) 403.
%
\bibitem{Dur}
   B. Durhuus, Nucl. Phys. {\bf B426} (1994) 203.
%
\bibitem{JK}
   J. Jurkiewicz and A. Krzywicki, Phys. Lett. {\bf B392} (1997) 291.
%
\bibitem{ABKW}
   J. Ambj\o rn, D.V. Boulatov, N. Kawamoto and Y. Watabiki,
   Phys. Lett. {\bf B480} (2000) 319.
%




%
%
%
%
%


%

%
%
























%

\end{thebibliography}
\end{document}